\documentclass[a4paper,11pt]{article}
\pdfoutput=1

\usepackage[svgnames]{xcolor} 

\usepackage[printonlyused,withpage]{acronym} 
\usepackage{amsfonts}
\usepackage{anyfontsize}
\usepackage{bbm}
\usepackage{booktabs} 
\usepackage{braket}
\usepackage{cancel}
\usepackage{changepage}
\usepackage{colortbl}
\usepackage{enumitem}
\usepackage[mathscr]{euscript}
\usepackage{float}
\usepackage[T1]{fontenc} 
\usepackage{gensymb}
\usepackage{Packages/jheppub}
\usepackage{listings} 
\usepackage{lmodern} 
\usepackage{makecell}
\usepackage{mathtools}
\usepackage[framemethod=tikz]{mdframed} 
\usepackage{microtype} 
\usepackage{multirow}
\usepackage{physics}
\usepackage{relsize}
\usepackage{rotating} 
\usepackage{sansmath}
\usepackage{simplewick}
\usepackage{slashed}
\usepackage{stmaryrd}
\usepackage{subfig}
\usepackage{tcolorbox}
\usepackage{theorem}
\usepackage{tikz}
\usepackage{tikz-feynman,contour}
\usepackage[normalem]{ulem}
\usepackage{xspace} 
\usepackage{yfonts}
\usepackage[vcentermath]{youngtab}

\tcbuselibrary{theorems}
\usetikzlibrary{decorations.markings}

\def\d{{\mathrm{d}}} 

\def\ii{{\text{i}}}

\def\doot{{\boldsymbol{\hspace{0.1em} \cdot\hspace{0.1em}}}}

\makeatletter
\newcommand*{\transpose}{%
  {\mathpalette\@transpose{}}%
}
\newcommand*{\@transpose}[2]{%
  \raisebox{\depth}{$\m@th#1\intercal$}%
}
\makeatother

\newtcbox{\sln}{colback=Gainsboro,
colframe=Gainsboro}

\newcommand{\tv}[1]{\overset{{}_{\,\scalebox{0.55}{$\shortrightarrow$}}}{#1}}

\newcommand{\bt}[1]{{\sansmath{\boldsymbol{#1}}}}

\newcommand{\overbar}[1]{\mkern 2mu\overline{\mkern-4mu#1\mkern-4mu}\mkern 2mu}

\tikzset{snake it/.style={decorate, decoration={snake,amplitude=10mm}}}

\tikzset{/pgf/decoration/.cd,
    number of sines/.initial=10,
    angle step/.initial=20,
}

\newdimen\tmpdimen

\pgfdeclaredecoration{full sines}{initial}
{
    \state{initial}[
        width=+0pt,
        next state=move,
        persistent precomputation={
            \pgfmathparse{\pgfkeysvalueof{/pgf/decoration/angle step}}%
            \let\anglestep=\pgfmathresult%
            \let\currentangle=\pgfmathresult%
            \pgfmathsetlengthmacro{\pointsperanglestep}%
                {(\pgfdecoratedremainingdistance/\pgfkeysvalueof{/pgf/decoration/number of sines})/360*\anglestep}%
        }] {}
    \state{move}[width=+\pointsperanglestep, next state=draw]{
        \pgfpathmoveto{\pgfpointorigin}
    }
    \state{draw}[width=+\pointsperanglestep, switch if less than=1.25*\pointsperanglestep to final, 
        persistent postcomputation={
        \pgfmathparse{mod(\currentangle+\anglestep, 360)}%
        \let\currentangle=\pgfmathresult%
    }]{%
        \pgfmathsin{+\currentangle}%
        \tmpdimen=\pgfdecorationsegmentamplitude%
        \tmpdimen=\pgfmathresult\tmpdimen%
        \divide\tmpdimen by2\relax%
        \pgfpathlineto{\pgfqpoint{0pt}{\tmpdimen}}%
    }
    \state{final}{
        \ifdim\pgfdecoratedremainingdistance>0pt\relax
            \pgfpathlineto{\pgfpointdecoratedpathlast}
        \fi
   }
}

\pgfdeclaredecoration{completely sines}{initial}
{
    \state{initial}[
        width=+0pt,
        next state=upsine,
        persistent precomputation={\pgfmathsetmacro\matchinglength{
            \pgfdecoratedinputsegmentlength / int(\pgfdecoratedinputsegmentlength/\pgfdecorationsegmentlength)}
            \setlength{\pgfdecorationsegmentlength}{\matchinglength pt}
        }] {}
    \state{upsine}[width=\pgfdecorationsegmentlength,next state=downsine]{
        \pgfpathsine{\pgfpoint{0.25\pgfdecorationsegmentlength}{0.5\pgfdecorationsegmentamplitude}}
        \pgfpathcosine{\pgfpoint{0.25\pgfdecorationsegmentlength}{-0.5\pgfdecorationsegmentamplitude}}
    }
    \state{downsine}[width=\pgfdecorationsegmentlength,next state=upsine]{
        \pgfpathsine{\pgfpoint{0.25\pgfdecorationsegmentlength}{-0.5\pgfdecorationsegmentamplitude}}
        \pgfpathcosine{\pgfpoint{0.25\pgfdecorationsegmentlength}{0.5\pgfdecorationsegmentamplitude}}
}
    \state{final}{}
}

\tikzfeynmanset{ with arrow/.style = {
   decoration={
     markings,
     mark=at position 0.5
          with {\arrow[xshift=1.9mm]{Latex[width=2.75mm,length=3mm]}}
     },
   postaction=decorate}
}

\tikzfeynmanset{ with glarrow/.style = {
   decoration={
     markings,
     mark=at position 0.5
          with {\arrow[white,xshift=3mm]{Latex[width=4.125mm,length=4.5mm]}}
     },
   postaction=decorate}
}

\tikzfeynmanset{ with outarrow/.style = {
   decoration={
     markings,
     mark=at position 0.5
          with {\arrow[xshift=3.35mm]{Latex[width=4.58333mm,length=5mm]}}
     },
   postaction=decorate}
}

\tikzfeynmanset{ bigphoton/.style={
    /tikz/draw=none,
    /tikz/decoration={name=none},
    /tikz/postaction={
      /tikz/draw,
      /tikz/decoration={
        completely sines,
        segment length=3mm,
        amplitude=2.5mm,
      },
      /tikz/decorate=true,
    },
  },
}

\tikzfeynmanset{ roundbigphoton/.style={
    /tikz/draw=none,
    /tikz/decoration={name=none},
    /tikz/postaction={
      /tikz/draw,
      /tikz/decoration={
        full sines,
        segment length=3mm,
        amplitude=1.25mm,
      },
      /tikz/decorate=true,
    },
  },
}

\tikzset{ mega thick/.style= {line width = 3.4pt}
}

\hypersetup{colorlinks, breaklinks, linkcolor=black,citecolor=black,filecolor=black,urlcolor=black} 

\makeatletter
\renewcommand{\fnum@figure}{\textsc{\figurename~\thefigure}} 
\makeatother

\lstnewenvironment{pseudoc}{\lstset{frame=lines,language=C,mathescape=true}}{}
\lstnewenvironment{logs}{\lstset{frame=lines,basicstyle=\footnotesize\ttfamily,numbers=none}}{}
\lstnewenvironment{cc}{\lstset{frame=lines,language=C}}{}
\lstnewenvironment{c64}{\lstset{backgroundcolor=\color{c64},basewidth=0.65em,basicstyle=\commodoreface\color{c64light},numbers=none,framerule=10pt,rulecolor=\color{c64light},frame=tb,framexbottommargin=30pt}}{}
\lstnewenvironment{html}{\lstset{frame=lines,language=html,numbers=none}}{}
\lstnewenvironment{pseudo}{\lstset{frame=lines,mathescape=true,morekeywords={learn_string_domain, save_model}}}{}
\lstnewenvironment{pseudoctiny}{\lstset{language=C,mathescape=true,basicstyle=\tiny\sffamily}}{}
\lstnewenvironment{cctiny}{\lstset{language=C,basicstyle=\tiny\sffamily}}{}
\lstnewenvironment{pseudotiny}{\lstset{mathescape=true,basicstyle=\tiny\sffamily}}{}

\title{Distinctive signals of frustrated dark matter}

\author[a,b]{Linda M. Carpenter,}
\author[a,b]{Taylor Murphy,}
\affiliation[a]{Department of Physics, The Ohio State University\\
191 W. Woodruff Ave., Columbus, OH 43210, U.S.A.}
\affiliation[b]{Center for Cosmology and Astroparticle Physics (CCAPP), The Ohio State University\\
191 West Woodruff Avenue, Columbus, OH 43210, U.S.A.}

\author[c]{and Tim M. P. Tait}
\affiliation[c]{Department of Physics and Astronomy, University of California, Irvine\\ Irvine, CA 92697, U.S.A.}

\emailAdd{lmc@physics.osu.edu}
\emailAdd{murphy.1573@osu.edu}
\emailAdd{ttait@uci.edu}

\date{\today}

\abstract{\begin{abstract}

We study a renormalizable model of Dirac fermion dark matter (DM) that communicates with the Standard Model (SM) through a pair of mediators --- one scalar, one fermion --- in the representation $(\boldsymbol{6},\boldsymbol{1}, \tfrac{4}{3})$ of the SM gauge group $\mathrm{SU}(3)_{\text{c}} \times \mathrm{SU}(2)_{\text{L}} \times \mathrm{U}(1)_Y$. While such assignments preclude direct coupling of the dark matter to the Standard Model at tree level, we examine the many effective operators generated at one-loop order when the mediators are heavy, and find that they are often phenomenologically relevant. We reinterpret dijet and pair-produced resonance and $\text{jets} + E_{\text{T}}^{\text{miss}}$ searches at the Large Hadron Collider (LHC) in order to constrain the mediator sector, and we examine an array of DM constraints ranging from the observed relic density $\Omega_{\chi} h^2_{\text{Planck}}$ to indirect and direct searches for dark matter. Tree-level annihilation, available for DM masses starting at the TeV scale, is required in order to produce $\Omega_{\chi} h^2_{\text{Planck}}$ through freeze-out, but loops --- led by the dimension-five DM magnetic dipole moment --- are nonetheless able to produce signals large enough to be constrained, particularly by the XENON1T experiment. In some benchmarks, we find a fair amount of parameter space left open by experiment and compatible with freeze-out. In other scenarios, however, the open space is quite small, suggesting a need for further model-building and/or non-standard cosmologies.

\end{abstract}}

\begin{document}

\maketitle

\section{Introduction}
\label{s1}

Despite a wealth of evidence that it comprises around eighty-five percent of the matter in the Universe \cite{Bertone:2004pz,Planck_2016,RD_2020}, 
the nature of dark matter (DM) remains unknown. 
The hypothesis that dark matter is composed of particles has generated an enormous body of work in an attempt to characterize and discover those particles
\cite{Bertone:2018krk}. 
Experimentally, particle dark matter has been targeted in a variety of ways, ranging from searches for DM produced invisibly at 
terrestrial particle colliders \cite{albert2017recommendations} to \emph{direct} searches for DM interacting with nuclei \cite{LUX_2017,X1T_2017,PhysRevLett.118.251301} 
and \emph{indirect} searches for visible signatures of cosmic DM annihilation \cite{IC_2013,PhysRevLett.117.091103,LAT_2017}. 
In the absence of any definitive signal of physics beyond the Standard Model (bSM), this large and multifaceted experimental effort has 
excluded large swaths of parameter space for dark matter candidates in many popular frameworks.

For their part, theorists have offered a plethora of extensions of the Standard Model (SM) that include one or more DM candidates. 
Some of these models, including supersymmetric models, are fairly complete --- at least to scales far above the weak scale --- 
but often suffer from very high-dimensional parameter spaces. 
Effective field theories (EFTs), by ignoring microscopic details, offer a more model-independent approach
\cite{Cao:2009uw,Beltran:2008xg,Beltran:2010ww,Goodman:2010ku}. 
Models of this class can parametrize a variety of DM couplings to the SM generated by heavy fields that are integrated out,
leaving behind a low-energy description in terms of a universal set of non-renormalizable interactions.
However, the large collision energies of the LHC imply that for many theories of interest the mediators are directly accessible,
requiring that they be directly included. An increasingly popular compromise is furnished by 
\emph{simplified models} \cite{LHCNewPhysicsWorkingGroup:2011mji}, 
which specify the microscopic interactions relevant to the DM candidate(s) while 
remaining agnostic about any other bSM physics, including the full ultraviolet (UV) completion. 
Such models trade the versatility of EFTs for superior detail and 
applicability up to higher energy scales. As there naturally exists a large array of gauge-invariant and renormalizable UV completions, 
even after decades of work there remains a vast landscape ripe for theoretical and experimental exploration.

We examine a family of simplified models that generate a wide variety of couplings between dark matter and the visible sector at one-loop order. In these models, the dark matter is a Dirac fermion $\chi$ transforming as a singlet under the SM gauge group, and all mediator fields coupling both to $\chi$ and to SM fields carry SM gauge charges that preclude renormalizable gauge-invariant interactions between the dark matter and any SM fermion. These models therefore require a \emph{pair} of mediators, one scalar $\varphi$ and one Dirac fermion $\psi$, in order to construct gauge- and Lorentz-invariant interactions between the dark matter and the Standard Model. For a variety of gauge assignments, one or both of the mediators may have renormalizable interactions with the SM, but the dark matter itself interacts directly with the SM only at loop level. This simple framework captures a general class of possibilities in which the mediator sector is charged under the SM, but the
interactions of the dark matter are \emph{frustrated} in the sense that the specific mediator assignments preclude its tree level interaction
with the SM. It is a flexible framework, emblematic of a situation that one can easily imagine descending from
a more fundamental UV theory, and produces phenomenology that is rich (able to be explored using the tools of astrophysics and 
elementary particle physics) and highly sensitive to 
the details determining how each mediator interacts with the SM.
In this work, we consider one particular renormalizable realization of this model framework 
wherein the mediators are $\mathrm{SU}(3)_{\text{c}}$ sextets and (only) the scalar couples to SM quarks. 
We identify a multitude of important signatures of this model and explore its parameter space in light of both terrestrial and astrophysical experiments.

This paper is organized as follows. In \hyperref[s2]{Section 2}, we introduce the model, 
giving particular attention to the hypercharged color-sextet mediator sector. Our phenomenological investigation begins in \hyperref[s3]{Section 3} 
with a survey of LHC searches for dijet resonances, both singly and pair produced, and for jets accompanied by missing 
transverse energy ($E_{\text{T}}^{\text{miss}}$); we chiefly use these to find open parameter space for the mediators, though some constraints 
can already be imposed on the dark matter here. After choosing several benchmark points in the mediator parameter space, we explore in \hyperref[s4]{Section 4} the characteristic sizes and physical consequences of the many effective operators generated by integrating out the heavy mediators. 
We finally turn to the dark matter in \hyperref[s5]{Section 5}, exploring cosmological and astrophysical constraints 
to see what parameter space satisfies the traditional dark matter observables. We draw conclusions and consider future work in \hyperref[s6]{Section 6}.
\section{Dark matter with bipartite mediators}
\label{s2}

We consider a family of models featuring Dirac dark matter, which is a Standard Model singlet,
coupled to a pair of mediators, each carrying color and hypercharge, schematically of the form
\begin{align*}
    \text{SM}\ \longleftrightarrow\ \text{mediators}\ \begin{Bmatrix}
    \text{$\varphi$ (scalar)}\\
    \text{$\psi$ (Dirac)}\end{Bmatrix}\ \longleftrightarrow\ \text{DM $\chi$},
\end{align*}
where the mediators carry both color and hypercharge. While this family of models includes realizations with mediators carrying $\mathrm{SU}(2)_{\text{L}}$ (weak isospin), we set these aside and consider weak singlets only. As discussed below, we specialize to the case of mediators transforming 
in the six-dimensional (sextet, $\boldsymbol{6}$) representation of $\mathrm{SU}(3)_{\text{c}}$. With this
gauge assignment, two mediators are necessary in order to construct renormalizable and Lorentz-invariant interactions with the singlet fermion dark matter. The choice of their weak hypercharge $Y$, meanwhile, crucially determines the allowed decays and experimental signatures; we also discuss this further below. 
More concretely,
\begin{align*}
    \mathcal{L} = \mathcal{L}_{\text{SM}} + \mathcal{L}_{\text{med}} + \mathcal{L}_{\chi},
\end{align*}
where $\mathcal{L}_{\text{SM}}$ is the Standard Model Lagrangian density;
\begin{align}\label{e3}
    \mathcal{L}_{\text{med}} = (D_{\mu}\varphi)^{\dagger s} (D^{\mu}\varphi)_s - m_{\varphi}^2 \varphi^{\dagger s} \varphi_s + \bar{\psi}^s(\ii \slashed{D} - m_{\psi})\psi_s + \mathcal{L}_{\text{decay}}
\end{align}
governs the mediators; and
\begin{align}\label{e4}
 \mathcal{L}_{\chi} = \bar{\chi}(\ii \slashed{\partial} - m_{\chi})\chi + y_{\chi}(\varphi^{\dagger s} \bar{\chi}\psi_s + \text{H.c.})
\end{align}
describes the dark matter and its interaction with the mediator pair. 
In \eqref{e3}, the gauge-covariant derivative $D^{\mu}$ acting on a generic field $O$ of 
hypercharge $Y$ and $\mathrm{SU}(3)_{\text{c}}$ representation r with indices $\{s,t\}$ is given 
by\footnote{Here $g_1$ and $g_3$ are the weak hypercharge and strong couplings, $g^{\mu}$ is a gluon field, 
and $[\bt{t}_{\text{r}}^a]_s^{\ \, t}$, with $a \in \{1,\dots,8\}$ and $\{s,t\} \in \{1,\dots,N_{\text{r}}\}$, 
are the generators of the $N_{\text{r}}$-dimensional representation r of $\mathrm{SU}(3)$. 
We normalize weak hypercharge so that the Gell-Mann--Nishijima relation is $Q = t^3 + Y$.}
\begin{align}\label{e5}
(D^{\mu}O)_s = [D^{\mu}]_s^{\ \, t} O_t = [(\partial^{\mu} - \ii g_1 Y B^{\mu})\delta_s^{\ \, t} - \ii g_3 [\bt{t}^a_{\text{r}}]_s^{\ \, t} g^{\mu}_a] O_t.
\end{align}
In order to guarantee that $\chi$ is the only stable DM particle, we impose a $\mathbb{Z}_2$ symmetry under which $\chi$ 
and \emph{one} of the mediators is odd and everything else is even, and we insist that the $\mathbb{Z}_2$-odd mediator be heavier than $\chi$.

The form of $\mathcal{L}_{\text{decay}}$, which determines how the mediators decay into SM particles, depends on the 
mediators' $\mathrm{SU}(3)_{\text{c}} \times \mathrm{U}(1)_Y$ representations and $\mathbb{Z}_2$ parities. 
We recently undertook a comprehensive study of color-sextet Dirac fermion and scalar interactions with Standard Model 
fields \cite{sextet_catalog} in which we identified many Lorentz- and gauge-invariant operators of mass dimension seven and below 
for both sextet species. 
Many of these are non-renormalizable operators that are straightforwardly generated from 
minimal ultraviolet completions and would be worthy candidates for the present investigation. 
But there is only one dimension-four operator; namely,
\begin{align}\label{L6}
\mathcal{L}_{\text{decay}} = \lambda_{IJ} \bt{K}_s{}^{ij}\, \varphi^{\dagger s} \overbar{q^{\text{c}}_{\text{R}}}_{Ii} q_{\text{R}Jj} + \text{H.c.}\ \ \ \text{with}\ \ \ q \in \{u,d\},
\end{align}
which couples a color-sextet scalar to a pair of right-chiral quarks, and has been previously studied in the 
literature \cite{Shu:2009xf,Han_2010,Han_2010_2}. 
Its size is controlled by couplings $\lambda_{IJ}$ in generation space ($I,J \in \{1,2,3\}$), and gauge invariance is ensured by the 
presence of the \emph{generalized Clebsch-Gordan coefficients} $\bt{K}_s{}^{ij}$, 
where now (and going forward) $s \in \{1,\dots,6\}$ is an $\mathrm{SU}(3)_{\text{c}}$ sextet index 
and $i,j \in \{1,2,3\}$ are the quark color ($\mathrm{SU}(3)_{\text{c}}$ fundamental) 
indices.\footnote{The Hermitian conjugates of these objects are denoted by $\bt{\bar{K}}^s{}_{ij}$. Compendia of technical details about these group-theoretical objects are available \cite{Han_2010,sextet_catalog}.} 
It is important to note that, in a model with a single type of sextet scalar, only one kind of coupling --- $uu, ud$, or $dd$ --- exists, as determined by
the sextet hypercharge assignment.

The operator \eqref{L6} offers an attractive building block to complete the mediator sector by introducing renormalizable 
mediator couplings to quarks, which dramatically enriches the phenomenology. We thus focus on models with a $\mathbb{Z}_2$-odd 
sextet fermion and a $\mathbb{Z}_2$-even sextet scalar coupled to quarks by some variant of \eqref{L6}. 
For definiteness, we choose to study mediators coupling to up-type quarks and therefore assign weak hypercharge $Y = 4/3$ to our sextet mediators. The quantum numbers of all novel fields in this particular model are listed in \hyperref[quantumNumTable]{Table 1}.
In the interest of simplicity, we restrict the couplings $\lambda_{IJ}$ to be real,
but we remain agnostic at this stage about the specific texture of $\lambda_{IJ}$. We discuss various schemes in our phenomenological study below. 

\renewcommand\arraystretch{1.4}
\begin{table}
    \centering
    \begin{tabular}{|c|c||c|c|}
\toprule
\hline
Field & Description & $\mathrm{SU}(3)_{\text{c}} \times \mathrm{SU}(2)_{\text{L}} \times \mathrm{U}(1)_Y$ representation & Couples to SM? \\
\hline
\hline
$\chi$ & Dark matter & $(\boldsymbol{1},\boldsymbol{1},0)$ & \\
\hline
$\varphi$ & Scalar mediator & \multirow{2}{*}[-0.0ex]{$(\boldsymbol{6},\boldsymbol{1},\tfrac{4}{3})$} & \checkmark \\
\cline{1-2}
\cline{4-4}
$\psi$ & Dirac mediator & & \\
\hline
\bottomrule
    \end{tabular}
    \caption{Novel field content in renormalizable model of Dirac dark matter investigated in this work. In this specific scheme, only $\varphi$ couples directly to SM fields at tree level.}
    \label{quantumNumTable}
\end{table}
\renewcommand\arraystretch{1.0}

We implement this model in version 2.3.43 of the \textsc{FeynRules} \cite{FR_OG,FR_2} package for \textsc{Mathematica}$^\copyright$\ version 12.0 \cite{Mathematica}, which we use to generate model files suitable for analytic computations and Monte Carlo simulations at leading order (LO) in 
the gauge couplings. In the former category, we employ version 3.11 of the \textsc{Mathematica} package \textsc{FeynArts} to construct a variety of amplitudes at one-loop order. We pass the resultant amplitudes to \textsc{FeynCalc} version 9.3.0 for symbolic evaluation 
including Passarino-Veltman reduction of tensor loop integrals and algebraic simplification \cite{MERTIG1991345,FC_9.0,FC_9.3.0}. 
In some cases we use \textsc{Package-X} version 2.1.1 \cite{Patel:2017px} via \textsc{FeynHelpers} version 1.3.0 \cite{feynhelpers} 
to aid in these tasks. 
For Monte Carlo event generation and to validate some analytic and semi-analytic results, we produce a model in the Universal FeynRules 
Output (UFO) format \cite{UFO} used as input for \textsc{MadGraph5\texttt{\textunderscore}aMC@NLO} 
(\textsc{MG5\texttt{\textunderscore}aMC}) version 3.3.1 \cite{MG5,MG5_EW_NLO} 
and some tools based on that framework (viz. \hyperref[s5]{Section 5}).
\section{Constraints on mediators}
\label{s3}

In this section, we analyze the allowed parameter space for the color-sextet mediators in light of constraints from the LHC. 
The color-sextet scalar is subject to important constraints both from LHC searches for color-charged resonances and 
low-energy constraints on neutral meson mixing associated with flavor-changing neutral currents. The color-sextet fermion
decays into the dark matter, resulting in collider signatures involving missing transverse momentum.

\subsection{Constraints on the color-sextet scalar}
\label{s3.1}

We begin with the scalar mediator, which (viz. \hyperref[s2]{Section 2}) is a close cousin to the $Y=4/3$ sextet diquark cataloged in \cite{sextet_catalog} --- albeit with different charge and couplings than the sextets highlighted in the latter section of that work --- and studied more intensely in \cite{Han_2010}.
The purpose of this discussion, relative to those works, is to summarize indirect constraints on such sextets from searches for flavor-changing neutral currents, to significantly update the direct limits from searches for light dijets by leveraging the LHC Run 2 dataset, and to contrast those new bounds with limits from searches for dijet pairs, which naturally offer a complementary probe of these scalars.

\subsubsection{\emph{D}$^{\text{0}}$-$\boldsymbol{\bar{\text{\emph{D}}}}{}^{\text{0}}$ mixing}

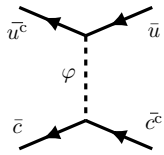
\begin{figure}[t]\label{FCNCs}
\begin{align*}
    \scalebox{0.75}{\begin{tikzpicture}[baseline={([yshift=-.5ex]current bounding box.center)},xshift=12cm]
\begin{feynman}[large]
\vertex (t1);
\vertex [below=1.5cm of t1] (t2);
\vertex [above left=0.5 cm and 1.1666cm of t1] (i1);
\vertex [right=1.75cm of t1] (f1);
\vertex [above right=0.5cm and 1.1666cm of t1] (p1);
\vertex [below right=0.5cm and 1.1666cm of t2] (p3);
\vertex [below left=0.5cm and 1.1666cm of t2] (i2);
\vertex [right=1.75cm of t2] (f2);
\vertex [below right=0.5 cm and 1.1666cm of f2] (p2);
\vertex [above right=0.5cm and 1.1666cm of f2] (p4);
\diagram* {
(t1) -- [ultra thick, fermion] (i1),
(t2) -- [ultra thick, fermion] (i2),
(t1) -- [ultra thick, scalar] (t2),
(p1) -- [ultra thick, fermion] (t1),
(p3) -- [ultra thick, fermion] (t2),
};
\end{feynman}
\node at (-1.2,0.1) {$\overbar{u^{\text{c}}}$};
\node at (-1.2,-1.6) {$\bar{c}$};
\node at (-0.3,-0.7) {$\varphi$};
\node at (1.2, 0.1) {$\bar{u}$};
\node at (1.2,-1.5) {$\overbar{c^{\text{c}}}$};
\end{tikzpicture}}
\end{align*}
\caption{Tree-level diagram mediated by a color-sextet scalar contributing to $D^0$-$\bar{D}{}^0$ mixing. (Diagram is drawn with external color-conjugate quarks to show correct flow of fermion number; this process is equivalent to $u\bar{c} \to \bar{u}c$.)}
\end{figure}

The tightest limits on the couplings of the sextet scalar to quarks,
$\lambda_{IJ}$ are from searches for flavor-changing neutral currents (FCNCs), 
which are observed to be (consistent with SM expectations) exceedingly small \cite{10.1093/ptep/ptaa104}, 
but can be enhanced by a color-sextet scalar coupling to up-type quarks \cite{PhysRevD.79.015017}. 
The most important potential FCNC enhancement is at tree level for $D^0$-$\bar{D}{}^0$ mixing as displayed in \hyperref[FCNCs]{Figure 1}.
There are additional box diagrams which contribute at next-to-leading order, but these diagrams vanish for diagonal sextet-quark couplings. The tree-level diagram, on the other hand, is proportional to $\lambda_{11}\lambda_{22}$. A recent analysis imposes a limit on this product of couplings \cite{PhysRevD.87.115019}:
\begin{align}\label{fcnclim}
(\lambda_{11}\lambda_{22})^2 \leq 9.3 \times 10^{-7} \left(\frac{m_{\varphi}}{\text{TeV}}\right)^2.
\end{align}
This is an extraordinarily stringent constraint that applies to any scenario with non-vanishing $\lambda_{11}$ and $\lambda_{22}$. For example, this constraint forbids $\lambda_{11} = \lambda_{22} \geq 5.8 \times 10^{-4}$ for a $600\,\text{GeV}$ sextet scalar with democratic coupling to up and charm quarks, $\lambda_{11} = \lambda_{22}$. Couplings so small would render single scalar production unobservable and obviate constraints from dijet-resonance searches (see below). 

Larger couplings can be viable if one assumes they have appropriate flavor structure. For example, invoking
\emph{minimal flavor violation} (MFV) \cite{Chivukula:1987py,Hall:1990ac,mfv_2002} by promoting the color-sextet fields to a set of flavor bi-triplets 
under quark-flavor symmetry group $\mathrm{SU}(3)_{Q_{\text{L}}} \times \mathrm{SU}(3)_{u_{\text{R}}} \times \mathrm{SU}(3)_{d_{\text{R}}}$ 
forbids tree-level $D^0$-$\bar{D}{}^0$ mixing. 
Another option is to consider a flavor texture with $\lambda_{22} = 0$; i.e., forbidding couplings to charm quarks, which
removes the FCNC constraint on $\lambda_{11}$.
For the present purposes, we investigate this second option, and
relegate a comprehensive investigation of an MFV scenario to future work.

\subsubsection{LHC searches}

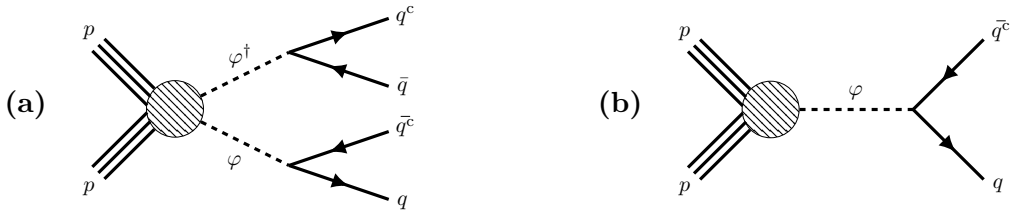
\begin{figure}\label{scalarLHC}
\begin{align*}
\textbf{(a)}\ \ \ \scalebox{0.75}{\begin{tikzpicture}[baseline={([yshift=-.5ex]current bounding box.center)},xshift=12cm]
\begin{feynman}[large]
\vertex (i1);
\vertex [right = 1.25cm of i1, blob] (i2){};
\vertex [below left = 0.15cm of i2] (q1);
\vertex [above left=0.46 cm of q1] (r1);
\vertex [below left =0.15cm of q1] (q2);
\vertex [above left = 0.39cm of q2] (r2);
\vertex [above left=0.15cm of i2] (l1);
\vertex [below left = 0.46cm of l1] (r3);
\vertex [above left=0.15cm of l1] (l2);
\vertex [below left=0.39cm of l2] (r4);
\vertex [above=0.2cm of i2] (q3);
\vertex [above=0.2cm of q3] (q4);
\vertex [above left=1.75cm of i2] (v3);
\vertex [below left=0.15cm of v3] (p1);
\vertex [below left =0.15cm of p1] (p2);
\vertex [below left=1.75cm of i2] (v4);
\vertex [above left=0.15cm of v4] (p3);
\vertex [above left =0.15cm of p3] (p4);
\vertex [above right=1 cm and 2cm of i2] (v1);
\vertex [above right=0.6 cm and 1.75cm of v1] (t1);
\vertex [below right=0.6cm and 1.75cm of v1] (t2);
\vertex [below right=1cm and 2cm of i2] (v2);
\vertex [above right=0.6 cm and 1.75cm of v2] (t3);
\vertex [below right=0.6 cm and 1.75cm of v2] (t4);
\diagram* {
(p1) -- [ultra thick] (r1),
(p2) -- [ultra thick] (r2),
(p3) -- [ultra thick] (r3),
(p4) -- [ultra thick] (r4),
(v3) -- [ultra thick] (i2),
(v4) -- [ultra thick] (i2),
(i2) -- [ultra thick, scalar ] (v1),
(i2) -- [ultra thick, scalar] (v2),
(t2) -- [ultra thick, fermion] (v1) -- [ultra thick, fermion] (t1),
(t3) -- [ultra thick, fermion] (v2) -- [ultra thick, fermion] (t4),
};
\end{feynman}
\node at (0.25,1.35) {$p$};
\node at (0.25,-1.4) {$p$};
\node at (2.8,-0.95) {$\varphi$};
\node at (2.9,0.95) {$\varphi^{\dagger}$};
\node at (5.8,1.65) {$q^{\text{c}}$};
\node at (5.75,0.4) {$\bar{q}$};
\node at (5.8,-0.3) {$\overbar{q^{\text{c}}}$};
\node at (5.75,-1.65) {$q$};
\end{tikzpicture}}\ \ \ \ \ \ \ \ \ \ \ \ \ \ \ \ \ \ \textbf{(b)}\ \ \ \scalebox{0.75}{\begin{tikzpicture}[baseline={([yshift=-.5ex]current bounding box.center)},xshift=12cm]
\begin{feynman}[large]
\vertex (i1);
\vertex [right = 1.25cm of i1, blob] (i2){};
\vertex [right=2.5cm of i2] (f1);
\vertex [above right=1.75cm of f1] (f2);
\vertex [below right=1.75cm of f1] (f3);
\vertex [below left = 0.15cm of i2] (q1);
\vertex [above left=0.46 cm of q1] (r1);
\vertex [below left =0.15cm of q1] (q2);
\vertex [above left = 0.39cm of q2] (r2);
\vertex [above left=0.15cm of i2] (l1);
\vertex [below left = 0.46cm of l1] (r3);
\vertex [above left=0.15cm of l1] (l2);
\vertex [below left=0.39cm of l2] (r4);
\vertex [above=0.2cm of i2] (q3);
\vertex [above=0.2cm of q3] (q4);
\vertex [above left=1.75cm of i2] (v3);
\vertex [below left=0.15cm of v3] (p1);
\vertex [below left =0.15cm of p1] (p2);
\vertex [below left=1.75cm of i2] (v4);
\vertex [above left=0.15cm of v4] (p3);
\vertex [above left =0.15cm of p3] (p4);
\vertex [above right=1 cm and 2cm of i2] (v1);
\vertex [above right=0.6 cm and 1.75cm of v1] (t1);
\vertex [below right=0.6cm and 1.75cm of v1] (t2);
\vertex [below right=1cm and 2cm of i2] (v2);
\vertex [above right=0.6 cm and 1.75cm of v2] (t3);
\vertex [below right=0.6 cm and 1.75cm of v2] (t4);
\diagram* {
(p1) -- [ultra thick] (r1),
(p2) -- [ultra thick] (r2),
(p3) -- [ultra thick] (r3),
(p4) -- [ultra thick] (r4),
(v3) -- [ultra thick] (i2),
(v4) -- [ultra thick] (i2),
(i2) -- [ultra thick, scalar ] (f1),
(f2) -- [ultra thick, fermion] (f1) -- [ultra thick, fermion] (f3),
};
\end{feynman}
\node at (0.25,1.35) {$p$};
\node at (0.25,-1.4) {$p$};
\node at (3.25,0.3) {$\varphi$};
\node at (5.8,1.4) {$\overbar{q^{\text{c}}}$};
\node at (5.75,-1.35) {$q$};
\end{tikzpicture}}
\end{align*}
\caption{\textbf{(a)} Pair and \textbf{(b)} single production of a color-sextet scalar at LHC. The scalar decays to an up-type quark pair $q_I q_I$ (assuming $Y = 4/3$ for the sextet and flavor-diagonal $\lambda_{IJ}$ in \eqref{L6}) if $m_{\varphi} > 2m_{q_I}$.}
\end{figure}

Color-sextet scalars can be copiously produced at hadron colliders both singly and in 
pairs \cite{Shu:2009xf,Han_2010,Han_2010_2,sextet_catalog} (see \hyperref[scalarLHC]{Figure 2}). The rate for pair production, 
mostly due to gluon fusion ($gg \to \varphi^{\dagger}\varphi$), is dominantly controlled by the gauge coupling, and 
is thus independent of the sextets' hypercharge $Y$. Single production, on the other hand, proceeds purely via 
quark-antiquark ($q\bar{q}$) annihilation due to \eqref{L6}, and is thus more model-dependent. 
Since we choose $Y = 4/3$, only up-type quark annihilation contributes. 
Moreover, our simplifying choice of a flavor-diagonal sextet-quark coupling $\lambda_{IJ}$, $I \in \{1,2,3\}$, 
restricts us further to $u\bar{u}$ and $c\bar{c}$ initial states. Once produced and under the assumption that $\lambda_{22} = 0$,
they typically decay into a pair of up- or top-quarks.
We therefore focus on the four processes
\begin{align*}
    pp \to \varphi^{\dagger}\varphi \to uu\bar{u}\bar{u}\ (tt\bar{t}\bar{t})\ \ \ \text{and}\ \ \ p p \to \varphi \to uu\ (tt).
\end{align*}
Mixed processes such as $pp \to \varphi^{\dagger}\varphi \to u u \bar{t}\bar{t}$, are also possible, but have received less attention
from the LHC experimental collaborations. 

Pair production followed by decays to light quarks \cite{CMS_EXO_17_021,ATLAS:2017jnp}
is most stringently constrained by the CMS search CMS-EXO-17-021, which looks for pair-produced resonances decaying to pairs of light (non-top) quarks 
using $35.9\,\text{fb}^{-1}$ of $pp$ collisions at $\sqrt{s}=13\,\text{TeV}$ at the LHC \cite{CMS_EXO_17_021}. The search
is divided into two regimes of putative resonance mass: $m_{\text{res}} \in [80,400)\,\text{GeV}$, where the resonances are typically highly boosted, and their decay products
reconstruct as a single jet, and $m_{\text{res}} \in (400,1500]\,\text{GeV}$, for which all four final-state quarks are typically reconstructed separately.
No excess is found relative to Standard Model expectations, and CMS places bounds on a supersymmetric model in which pair-produced scalar top quarks (stops) 
each decay to a like-sign quark pair ($\tilde{t} \to \bar{q}\bar{q}'$) via the $R$-parity-violating coupling $\lambda''_{312}$. As the kinematic
structure of this process is identical to that of the sextet scalar pair production, we translate these limits 
by appropriately rescaling the cross section. 

\begin{figure}
    \centering
    \includegraphics[scale=0.7]{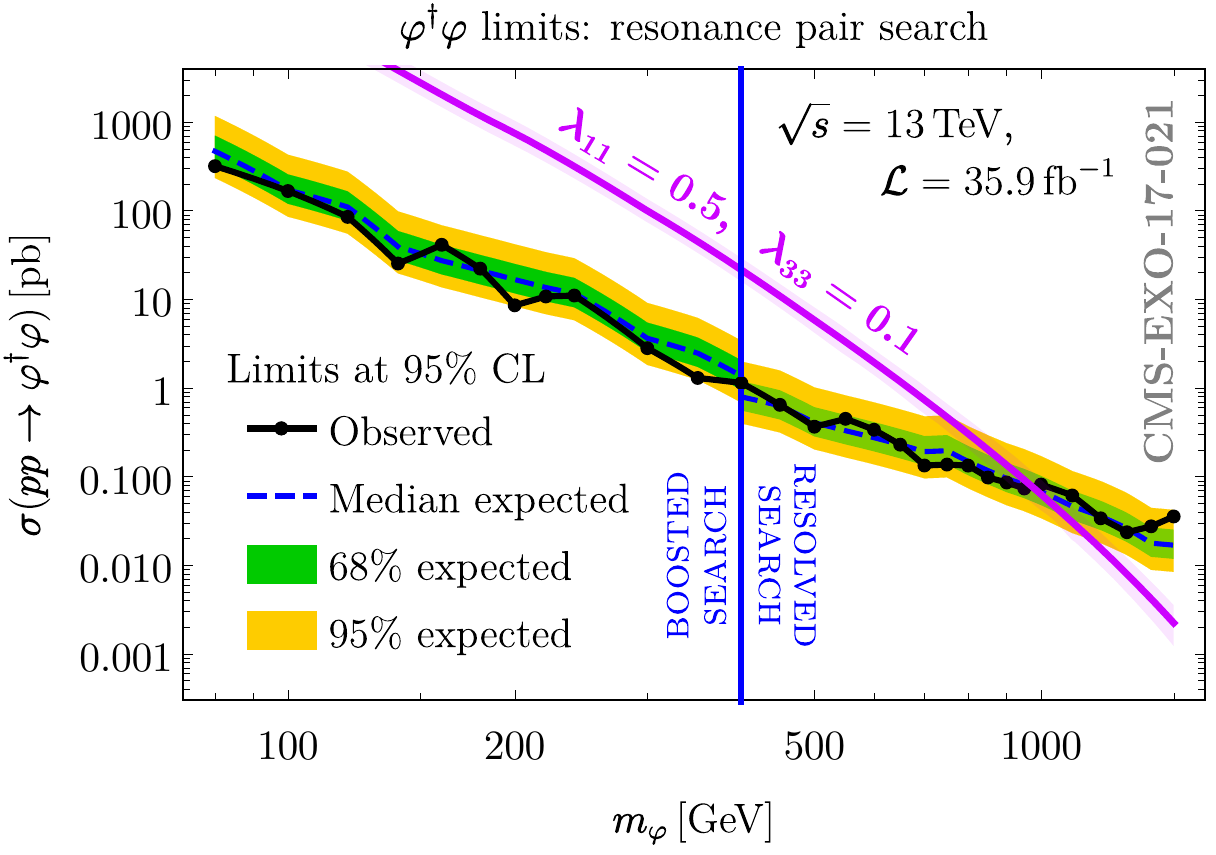}
    \caption{Limits on sextet scalar pair production from CMS-EXO-17-021, a search at $\sqrt{s}=13\,\text{TeV}$ for pair-produced resonances decaying to light quarks. 
    Displayed in purple is $\sigma(pp \to \varphi^{\dagger} \varphi \to uu \bar{u}\bar{u})$ for $\lambda_{11}=0.5$, $\lambda_{33}=0.1$.}
    \label{CMS4urecast1}
\end{figure}

\begin{figure}
    \centering
    \includegraphics[scale=0.65]{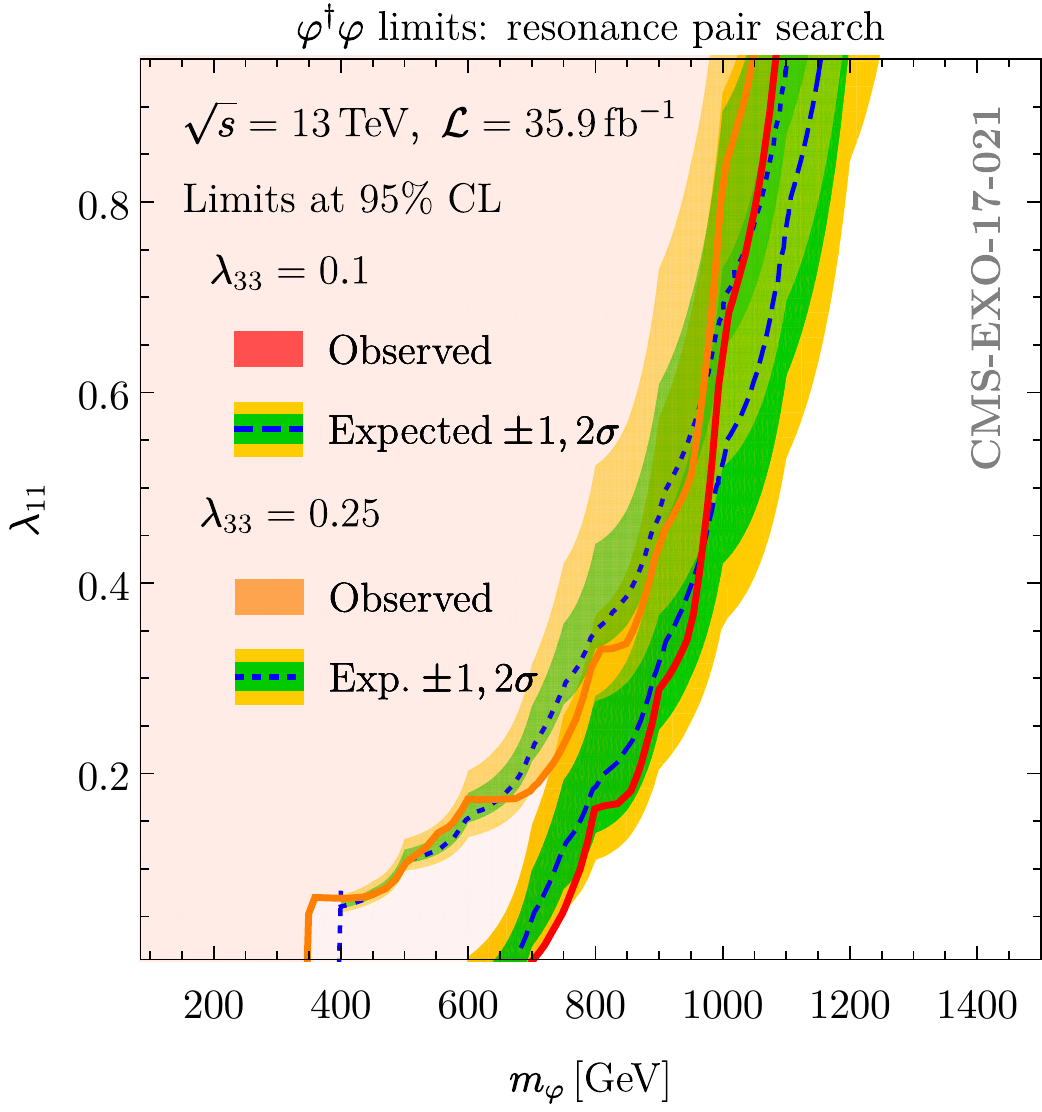}\hfill\includegraphics[scale=0.65]{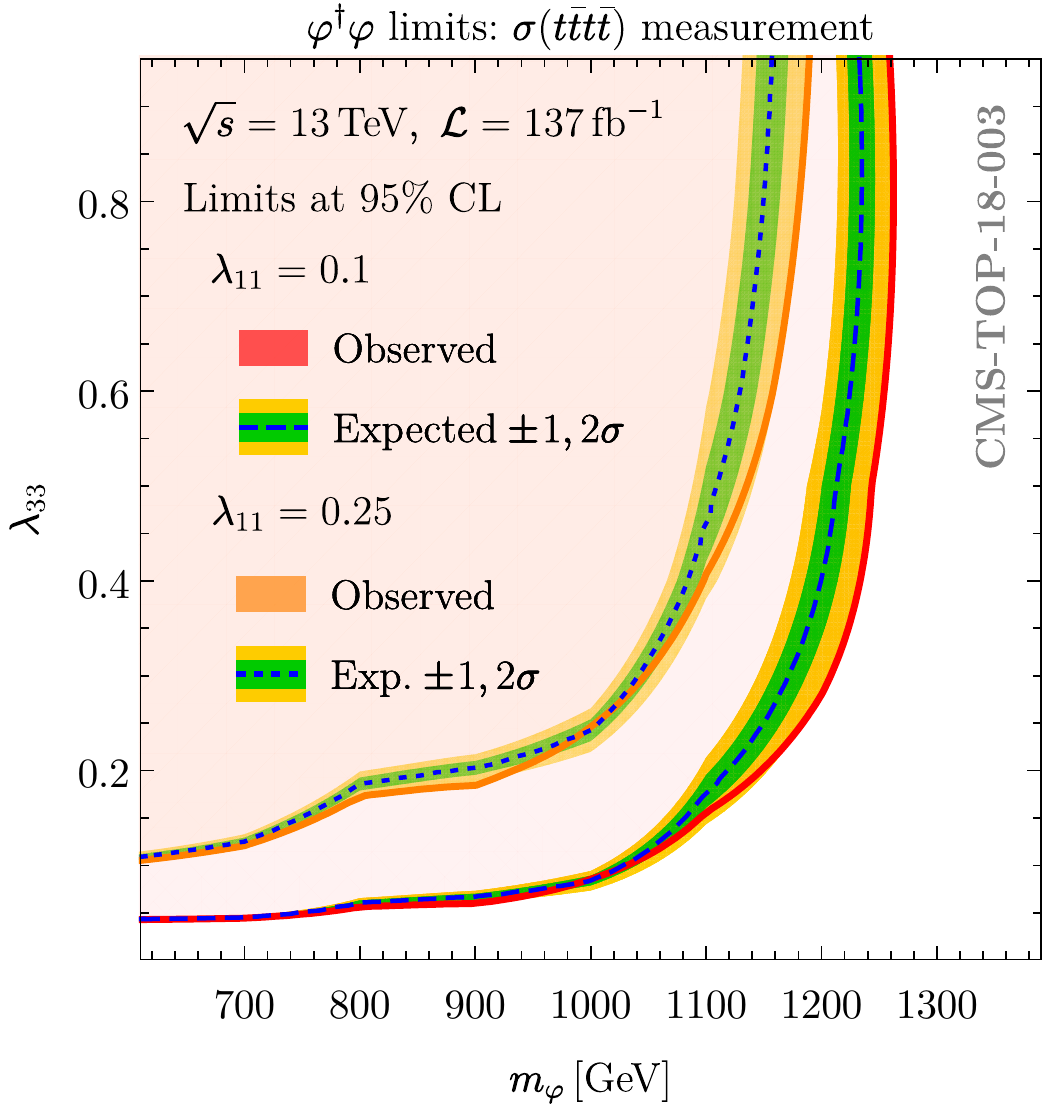}
    \caption{Limits on sextet scalar pair production from (left) CMS-EXO-17-021 and (right) CMS-TOP-18-003. 
    Left panel shows limits on $\lambda_{11}$ via $\sigma(pp \to \varphi^{\dagger}\varphi \to uu\bar{u}\bar{u})$ for two fixed values of $\lambda_{33}$; right panel displays the opposite via $\sigma(pp \to \varphi^{\dagger}\varphi \to tt\bar{t}\bar{t})$.}
    \label{CMS4qrecasts}
\end{figure}

The results of this simple reinterpretation are displayed in \hyperref[CMS4urecast1]{Figure 3}.
\hyperref[CMS4urecast1]{Figure 3} displays the leading order cross section for sextet scalar pair production followed by decays to up quarks, 
$\sigma(pp \to \varphi^{\dagger}\varphi \to uu\bar{u}\bar{u})$, for $\lambda_{11}=0.5$ and $\lambda_{33}=0.1$ using \textsc{MG5\texttt{\textunderscore}aMC}
with NNPDF\,2.3\,LO parton distribution functions \cite{nnpdf}, and renormalization and factorization scales set to the mass of the sextet scalar. 
For this benchmark, the sextet pair cross section is about an order of magnitude larger than the stop pair --- a consequence of the larger sextet color factor. We find an observed (expected) 95\% confidence level (CL) \cite{Read:2002cls} lower limit on $m_\varphi$ of 973 (981)\,\text{GeV}.
This limit has some dependence on both $\lambda_{11}$ and $\lambda_{33}$ through their impact on the total decay width and individual branching fractions.
To illustrate this, we show in the left panel of \hyperref[CMS4qrecasts]{Figure 4} the 95\% CL upper limits in the $(m_{\varphi},\lambda_{11})$ 
plane\footnote{We compute cross sections for $\lambda_{11} \in \{0.01,0.05,0.1,0.5,1\}$ in \textsc{MG5\texttt{\textunderscore}aMC}, 
and interpolate between these values.} for two representative choices of $\lambda_{33}$. Evident from the figure, the
upper bound on $\lambda_{11}$ relax as $\lambda_{33}$ (and hence the branching fraction $\text{BF}(\varphi \to tt)$) rises. 
The feature at $m_{\varphi} \approx 400\,\text{GeV}$ for $\lambda_{33}=0.25$ is due to the branching fraction to $uu$ approaching unity below the $tt$ threshold.
For masses above $\sim \!1$ TeV, essentially any perturbative value of $\lambda_{11}$ is allowed.

The right panel of \hyperref[CMS4qrecasts]{Figure 4} displays the allowed region of the $(m_{\varphi},\lambda_{33})$ plane derived from the analogous
CMS search CMS-TOP-18-003 for the production of four top quarks ($t\bar{t}t\bar{t}$) using $137\,\text{fb}^{-1}$ at $\sqrt{s}=13\,\text{TeV}$ \cite{CMS_TOP_18_003}
(a similar ATLAS search \cite{ATLAS:2020hpj} is currently less constraining). 
This search performs both a cut-based and a boosted decision tree (BDT) analysis on semi-leptonic (two or more leptons and jets) final states; 
the BDT analysis finds evidence for the target process with observed (expected) significance of 2.6 (2.7) standard deviations above background. 
A maximum-likelihood fit performed within the BDT analysis produces a measured cross section of $\sigma(pp \to t\bar{t}t\bar{t}) = 12.6^{+5.8}_{-5.2}\,\text{fb}$, which is consistent with the Standard Model prediction of $\sigma_{\text{SM}}(pp \to t\bar{t}t\bar{t}) = 12.0^{+2.2}_{-2.5}\,\text{fb}$ at next-to-leading order (NLO) \cite{SM4tNLO}, allowing CMS to place a
95\% CL upper limit on beyond-the-Standard Model contributions greater than about $10\,\text{fb}$.

Since none of the model frameworks considered in CMS-TOP-18-003 are direct analogs to $pp \to \varphi^{\dagger}\varphi \to tt\bar{t}\bar{t}$, we derive bounds
based on the limit on the inclusive cross section, using the reimplementation \cite{DVN/OFAE1G_2020} of the cut-based analysis of 
CMS-TOP-18-003 in \textsc{MadAnalysis\,5} (MA5) version 1.9.20 \cite{Conte_2013}, a framework designed to emulate LHC analyses for application to (in principle) any bSM theory \cite{Conte_2014,Conte_2018}. 
This implementation of CMS-TOP-18-003 is available on the \textsc{MadAnalysis\,5} Public Analysis Database (PAD) \cite{Dumont_2015}. 
We supply a set of eleven samples of $10^4$ $pp \to \varphi^{\dagger}\varphi \to tt\bar{t}\bar{t}$ 
events generated in \textsc{MG5\texttt{\textunderscore}aMC} for $m_{\varphi} \in [600, \dots, 1500]\,\text{GeV}$ in intervals of $100\,\text{GeV}$ 
(including parton shower matching and hadronization via \textsc{Pythia\,8} version 8.244 \cite{Pythia}) as input. 
These samples are passed by \textsc{MadAnalysis\,5} to \textsc{Delphes\,3} version 3.4.2 \cite{Delphes_OG,Delphes_3} and \textsc{FastJet} version 3.3.3 \cite{FJ}, 
which respectively model the response of the CMS detector and perform object 
reconstruction.\footnote{The public implementation of CMS-TOP-18-003 contains a \textsc{Delphes} card optimized for this analysis.}
MA5 computes the acceptances of the reconstructed events and determines the 95\% CL 
upper limit on the number of signal events, given the numbers of expected and observed background events. 
We map these limits onto the right panel of \hyperref[CMS4qrecasts]{Figure 4}, by computing a range of $\sigma(pp \to \varphi^{\dagger} \varphi \to tt\bar{t}\bar{t})$ 
in much the same way as described for pair-produced scalars decaying to up quarks. We find that above about $1200\,\text{GeV}$, there is essentially no bound on the coupling strengths.


\begin{figure}
    \centering
    \includegraphics[scale=0.7]{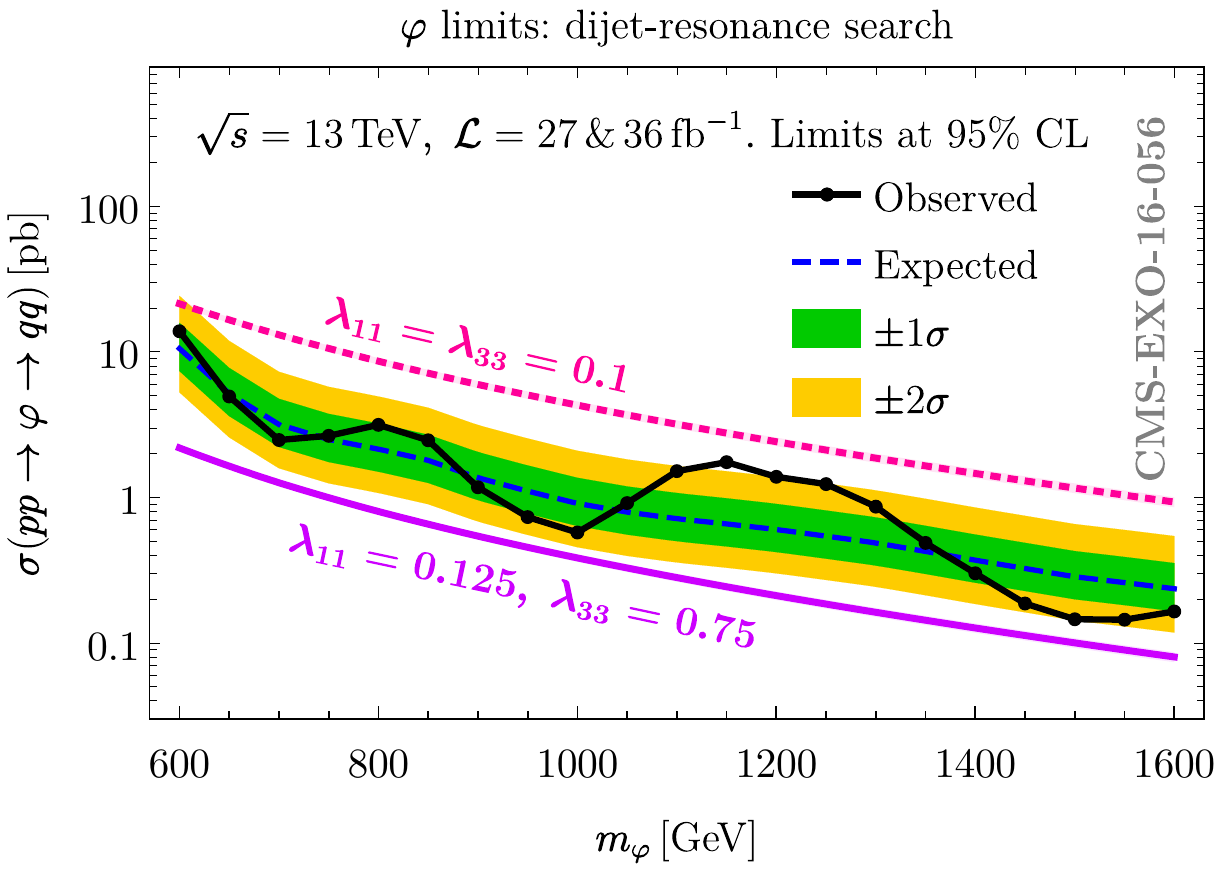}
    \caption{Limits on single scalar sextet production from CMS-EXO-16-056, a search at $\sqrt{s}=13\,\text{TeV}$ for dijet resonances decaying to light quarks. Displayed in purple (pink) is $\sigma(pp \to \varphi \to uu)$ for $\lambda_{11} = 0.125$, $\lambda_{33}=0.75$ ($\lambda_{11} = \lambda_{33} = 0.1$).}
    \label{CMSdijetrecast}
\end{figure}

Constraints on single production can be derived from searches for dijet 
resonances \cite{CMS_EXO_16_056,ATLAS:2019fgd,ATLAS:2020iwa}. The CMS search
CMS-EXO-16-056 uses up to $36\,\text{fb}^{-1}$ of $\sqrt{s}=13\,\text{TeV}$ data \cite{CMS_EXO_16_056}, with a low-mass regime, 
$m_{\text{res}} \in [0.6,1.6)\,\text{TeV}$, relevant for resonances decaying into pairs of light quarks which are reconstructed at the trigger level using data from the CMS calorimeter. 
No excess is observed, and 95\% CL upper limits on the production cross sections of narrow resonances are interpreted in the context of
a scalar diquark decaying to $qq$ \cite{HEWETT1989193} (among others). This simplified model has kinematic structure identical to that of single sextet scalar production,
and constraints can be inferred by rescaling the cross section. 
In \hyperref[CMSdijetrecast]{Figure 5}, we contrast the 95\% CL upper limit on the cross section for a dijet resonance from CMS-EXO-16-056 with two
predictions for $\sigma(pp \to \varphi \to uu)$ computed by \textsc{MG5\texttt{\textunderscore}aMC}
corresponding to two choices of $\lambda_{11}$ and $\lambda_{33}$ selected to bracket the CMS 
bound.\footnote{To be conservative, we do not include the $K$ factor $\sigma_{\text{NLO}}/\sigma_{\text{LO}} \simeq 1.2$ computed 
for the $u\bar{u}$ initial state and $m_{\varphi} \sim$~TeV in \cite{Han_2010}.} 
Unsurprisingly, the cross section is
highly sensitive to $\lambda_{11}$, while more flat with respect to the scalar mass, as single production probes the PDFs at significantly lower parton $x$. The role of $\lambda_{33}$, the coupling to top quarks, should not be overlooked: we see in \hyperref[CMSdijetrecast]{Figure 5} that the cross section with $\lambda_{11} = 0.125$ is smaller than for $\lambda_{11} = 0.1$ because the larger $\lambda_{33}$ gives $\varphi \to tt$ a much larger branching fraction. On that note, the top-quark process $pp \to \varphi \to tt$ can in principle be bounded by searches for resonances decaying into same-sign top pairs.
However, CERN-PH-EP-2012-020, a search by the ATLAS Collaboration, is the only search on record explicitly targeting this final state \cite{ATLAS:2012iws}. Because of its small luminosity and collision energy ($\mathcal{L} = 1.04\,\text{fb}^{-1}$ at $\sqrt{s}=7\,\text{TeV}$), it is considerably less impactful than the Run 2 searches, and does not impose additional constraints on the parameter
space. An updated analysis using similar techniques would likely provide useful bounds on $\lambda_{33}$.

\subsection{Constraints on the color-sextet fermion}
\label{s3.2}

The fermionic mediator $\psi$ is $\mathbb{Z}_2$-odd and heavier than the dark matter $\chi$. 
It can be pair-produced at colliders through its $\mathrm{SU}(3)_{\text{c}}$ gauge interaction, but is otherwise markedly different from the color-sextet fermions cataloged in \cite{sextet_catalog} since those couple directly to Standard Model fields and this sextet does not. Instead, it decays into $\chi$ plus two quarks (via an intermediate on- or off-shell $\varphi$),
leading to LHC signatures containing hard jets and missing transverse momentum ($E_{\text{T}}^{\text{miss}}$), as shown
schematically in \hyperref[fermionLHC]{Figure 6}.
\begin{figure}\label{fermionLHC}
\begin{align*}
    \scalebox{0.75}{\begin{tikzpicture}[baseline={([yshift=-.5ex]current bounding box.center)},xshift=12cm]
\begin{feynman}[large]
\vertex (i1);
\vertex [right = 1.25cm of i1, blob] (i2){};
\vertex [below left = 0.15cm of i2] (q1);
\vertex [above left=0.46 cm of q1] (r1);
\vertex [below left =0.15cm of q1] (q2);
\vertex [above left = 0.39cm of q2] (r2);
\vertex [above left=0.15cm of i2] (l1);
\vertex [below left = 0.46cm of l1] (r3);
\vertex [above left=0.15cm of l1] (l2);
\vertex [below left=0.39cm of l2] (r4);
\vertex [above=0.2cm of i2] (q3);
\vertex [above=0.2cm of q3] (q4);
\vertex [above left=1.75cm of i2] (v3);
\vertex [below left=0.15cm of v3] (p1);
\vertex [below left =0.15cm of p1] (p2);
\vertex [below left=1.75cm of i2] (v4);
\vertex [above left=0.15cm of v4] (p3);
\vertex [above left =0.15cm of p3] (p4);
\vertex [above right=1 cm and 2cm of i2] (v1);
\vertex [above right=0.6 cm and 1.75cm of v1] (t1);
\vertex [above right=0.5 cm and 1.75cm of t1] (f1);
\vertex [below right=0.5 cm and 1.75cm of t1] (f2);
\vertex [below right=0.6cm and 2.25cm of v1] (t2);
\vertex [below right=1cm and 2cm of i2] (v2);
\vertex [above right=0.6 cm and 2.25cm of v2] (t3);
\vertex [above right=0.5 cm and 1.75cm of t4] (f3);
\vertex [below right=0.5 cm and 1.75cm of t4] (f4);
\vertex [below right=0.6 cm and 1.75cm of v2] (t4);
\diagram* {
(p1) -- [ultra thick] (r1),
(p2) -- [ultra thick] (r2),
(p3) -- [ultra thick] (r3),
(p4) -- [ultra thick] (r4),
(v3) -- [ultra thick] (i2),
(v4) -- [ultra thick] (i2),
(v1) -- [ultra thick,color=blue, fermion] (i2),
(i2) -- [ultra thick,color=blue, fermion] (v2),
(t2) -- [ultra thick, color=blue,fermion] (v1),
(v1) -- [ultra thick, scalar] (t1),
(t4) -- [ultra thick, scalar] (v2),
(v2) -- [ultra thick, color=blue,fermion] (t3),
(f2) -- [ultra thick, fermion] (t1) -- [ultra thick, fermion] (f1),
(f3) -- [ultra thick, fermion] (t4) -- [ultra thick, fermion] (f4),
};
\end{feynman}
\node at (0.25,1.35) {$p$};
\node at (0.25,-1.4) {$p$};
\node at (2.8,-1) {$\color{blue}\psi$};
\node at (2.8,1) {$\color{blue}\bar{\psi}$};
\node at (4.6,-1.7) {$\varphi$};
\node at (6.25,-0.4) {$\color{blue}\chi$};
\node at (6.25,0.4) {$\color{blue}\bar{\chi}$};
\node at (4.7,1.675) {$\varphi^{\dagger}$};
\node at (7.55,2.15) {$q^{\text{c}}$};
\node at (7.5,1.1) {$\bar{q}$};
\node at (7.5,-2.15) {$q$};
\node at (7.55,-1.05) {$\overbar{q^{\text{c}}}$};
\end{tikzpicture}}
\end{align*}
\caption{Pair production of a $\mathbb{Z}_2$-odd color-sextet fermion (blue) at LHC. Each fermion can decay to the dark matter and (through the sextet scalar) two quarks, generating a jets + $E_{\text{T}}^{\text{miss}}$ signature, provided that $m_{\psi} > m_{\chi} + 2m_{q_I}$.}
\end{figure}
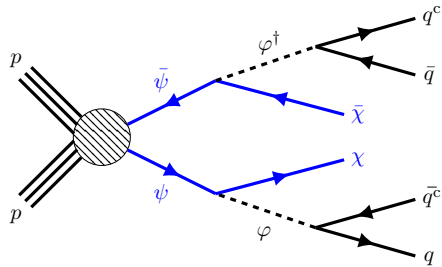This process is reminiscent of gluino pair production followed by decays to an 
neutralino and quarks via a (possibly off-shell) squark; e.g., $pp \to \tilde{g}\tilde{g} \to 2 \times( q\bar{q} + \tilde{\chi}^0)$, which produces jets + $E_{\text{T}}^{\text{miss}}$. 
There exist a pair of Run 2 searches targeting gluinos: ATLAS-CONF-2019-040 \cite{ATLAS-CONF-2019-040} (for light jets and $E_{\text{T}}^{\text{miss}}$ based on
$139\,\text{fb}^{-1}$)
and CMS-SUS-16-033 \cite{CMS-SUS-16-033} (for multijet events with $E_{\text{T}}^{\text{miss}}$ using $35.9\,\text{fb}^{-1}$). Neither search finds a significant excess, and both thus place 95\% CL limits on the allowed parameter space of the gluino and neutralino masses, under the assumption that the gluinos are pair-produced via the strong interaction. These searches can be reinterpreted to place bounds on our model parameter space and have been implemented in \textsc{MadAnalysis\,5}.

We simulate $10^4$ sextet fermion pair-production events followed by decays to quarks and dark matter, 
$pp \to \psi\bar{\psi} \to qq\chi + \bar{q}\bar{q}\bar{\chi}$, at LO, for each of about eighty points in the $(m_{\psi},m_{\chi})$ plane 
separated in $100\,\text{GeV}$ intervals, for each $q \in \{u,t\}$. 
Additional points have been added where finer detail is desired, such as at kinematic thresholds. 
We choose the sextet scalar-quark couplings $\lambda_{11} = 0.125$ and $\lambda_{33}=0.75$.\footnote{We fix $m_{\varphi} = 1.15\,\text{TeV}$, at the lower
bound of the what is allowed, and $y_{\chi}=1.0$. The inclusive cross sections are relatively insensitive to this choice, but the kinematic distributions are not.} 
Much as was done for the scalar mediator, we pass the Monte Carlo samples to 
\textsc{MadAnalysis\,5} to be analyzed in the context of the publicly available reimplementations \cite{Ambrogi_rec,DVN/NW3NPG_2021,DVN/GBDC91_2021} 
of ATLAS-CONF-2019-040 and CMS-SUS-16-033. 

\begin{figure}
    \centering
    \includegraphics[scale=0.7]{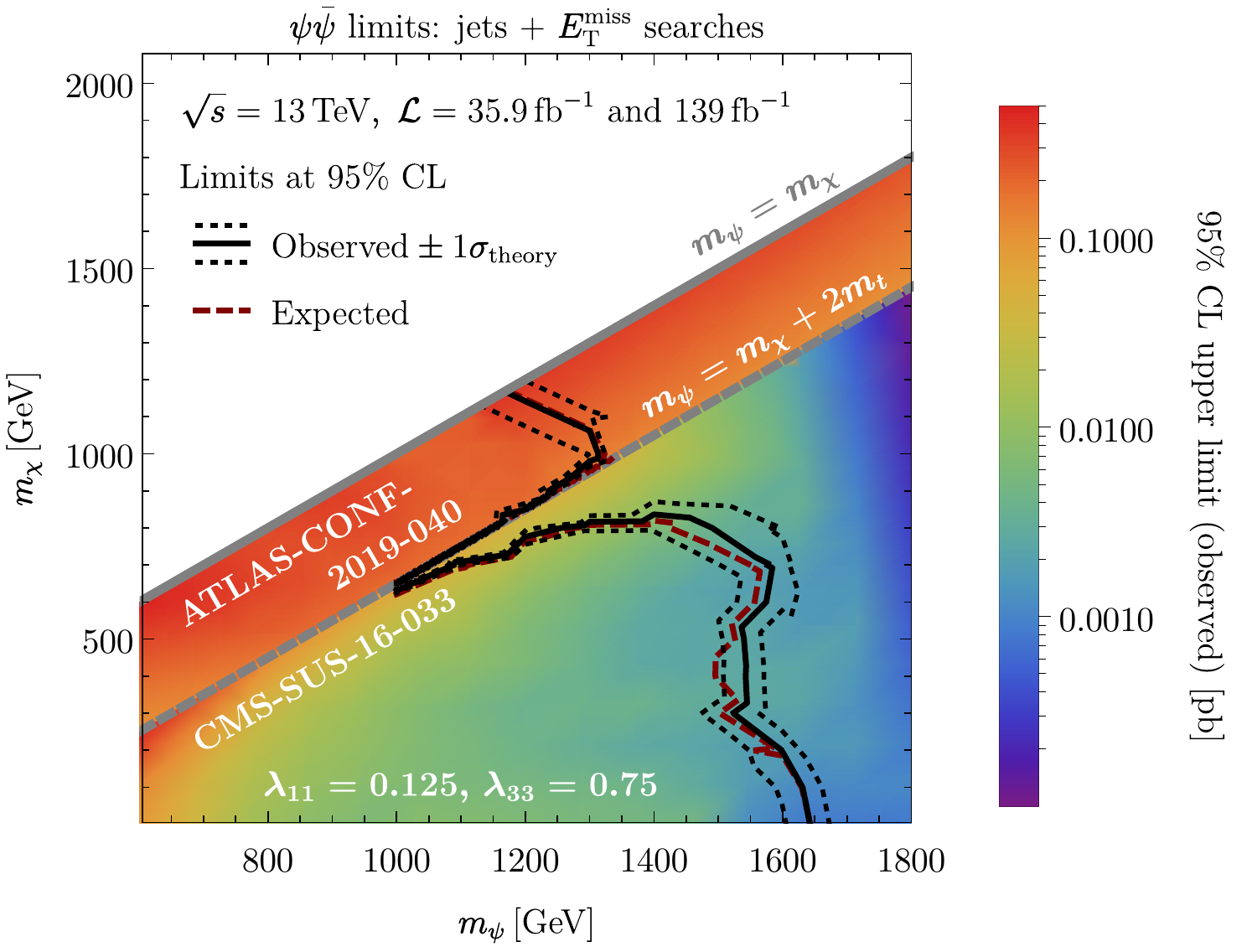}
    \caption{Limits on sextet fermion pair production from ATLAS-CONF-2019-040, a search at $\sqrt{s}=13\,\text{TeV}$ ($139\,\text{fb}^{-1}$) for gluinos in final states with jets and missing transverse momentum; and from CMS-SUS-16-033, a similar search ($35.9\,\text{fb}^{-1}$) for supersymmetry in multijet events with missing transverse momentum. First search targets light jets; second search encompasses heavy-flavor jets as seen from top quarks. Cross sections are computed for $m_{\varphi}= 1.15\,\text{TeV}$, $\lambda_{11}=0.125$, $\lambda_{33}=0.75$.}
    \label{jetsMETrecastfig}
\end{figure}

The results of these reinterpretations are displayed in \hyperref[jetsMETrecastfig]{Figure 7}, which shows the strongest observed and expected lower limits at 95\% CL 
from the pair of analyses at each point in the $(m_{\psi},m_{\chi})$ plane, in addition to the observed upper limit at 95\% CL on the cross section imposed by whichever analysis is most stringent at each point. The $\psi$ branching fraction to $tt + \chi$ rises quickly to almost unity once that channel is open, and thus ATLAS-CONF-2019-040, which is sensitive to light quarks,
provides the most stringent limit when decays to top quarks are not open, whereas CMS-SUS-16-033 is more restrictive when it is.
At the interface, there is a cleft near the $tt$ threshold where neither search can rule out a $\sim\! 1.0\,\text{TeV}$ sextet fermion, which could perhaps be filled in
by a search targeting a mixed scenario, $pp \to \psi \bar{\psi} \to uu\chi + \bar{t}\bar{t}\bar{\chi}$, etc. We observe that masses above $m_{\psi} = 1.65\,\text{TeV}$ are unconstrained, with even lighter choices of $m_\psi$ permitted for heavier dark matter.

While it is difficult to explore every scenario, these searches illustrate the allowed regions of parameter space for the mediators, and suggest benchmarks not already
excluded. After synthesizing the bounds from FCNC constraints and the LHC, we conclude that color-sextet scalars with masses 
$m_{\varphi} \sim 1.0\,\text{TeV}$ and couplings to quarks between $0.1$ and $1$ are allowed. Moderately heavier sextet scalars are viable for a wide variety of dark matter
masses. For the phenomenological investigation in the remainder of this work, we select the set of benchmarks displayed in \hyperref[benchmarktable]{Table 2}. These benchmarks are distinguished principally by the mass of the sextet fermion, which in turn controls the maximum DM mass. Our first benchmark, BP1, approaches the edge of the current LHC limits on both the sextet scalar and the fermion. The other two, BP2 and BP3, back away from the jets + $E_{\text{T}}^{\text{miss}}$ limits to produce a larger parameter space for the dark matter.

\renewcommand\arraystretch{1.4}
\begin{table}\label{benchmarktable}
    \begin{center}
        \begin{tabular}{|c || c | c | c || c | c|}
        \toprule
        \hline
 & $m_{\chi}$ range [GeV] & $m_{\varphi}$\,[GeV] & $m_{\psi}$\,[GeV] & $\lambda_{11}$ & $\lambda_{33}$\\
\hline
\hline
\rule{0pt}{2.5ex}BP1 & (0,\,1600) & \multirow{3}{*}[-0.25ex]{1150} & 1600 & \multirow{3}{*}[-0.25ex]{0.125} & \multirow{3}{*}[-0.25ex]{0.75}\\
\cline{1-2}
\cline{4-4}
\rule{0pt}{2.5ex}BP2 & (0,\,2000) &  & 2000 &  & \\
\cline{1-2}
\cline{4-4}
\rule{0pt}{2.5ex}BP3 & (0,\,5000) &  & 5000 &  & \\
\hline
\bottomrule
     \end{tabular}
    \end{center}
    \caption{Benchmark points for dark matter in models with color-sextet mediators consistent with LHC and FCNC constraints. The sextet-DM Yukawa coupling $y_{\chi}$ is permitted to vary in all benchmarks.}
\end{table}
\renewcommand\arraystretch{1}

\section{Dark matter loop interactions with the Standard Model}
\label{s4}

The gauge assignments of the mediators forbid direct tree-level coupling between the dark matter and the SM. Nevertheless, there are important interactions that arise at one
loop, which in the limit of heavy mediators can be described using effective field theory.
In this section we estimate the Wilson coefficients for potentially important DM-SM interactions and discuss the physical processes to which they contribute.

\subsection{Rayleigh operators}
\label{s4.1}

\begin{figure}\label{f1}
\begin{align*}
    \scalebox{0.75}{\begin{tikzpicture}[baseline={([yshift=-.5ex]current bounding box.center)},xshift=12cm]
\begin{feynman}[large]
\vertex [blob] (i1){};
\vertex [above left=0.75 cm and 1.75cm of i1] (d1);
\vertex [below left=0.75 cm and 1.75cm of i1] (d2);
\vertex [above right=0.75 cm and 1.75cm of i1] (v1);
\vertex [below right=0.75cm and 1.75cm of i1] (v2);
\diagram* {
(d1) -- [ultra thick, fermion] (i1),
(d2) -- [ultra thick] (i1),
(i1) -- [ultra thick, charged scalar] (d2),
(i1) -- [ultra thick, scalar] (d1),
(i1) -- [ultra thick, photon] (v1),
(i1) -- [ultra thick, photon] (v2),
};
\end{feynman}
\node at (-2,0.8) {$\chi$};
\node at (-2.0,-0.8) {$\bar{\chi}$};
\node at (2.06,0.75) {$V$};
\node at (2.05,-0.85) {$V$};
\end{tikzpicture}}\ \supset\ \scalebox{0.75}{\begin{tikzpicture}[baseline={([yshift=-.75ex]current bounding box.center)},xshift=12cm]
\begin{feynman}[large]
\vertex (i1);
\vertex [right = 1.25cm of i1] (i2);
\vertex [above left=1.5 cm of i2] (d1);
\vertex [below left=1.5cm of i2] (d2);
\vertex [right= 1.25cm of i2] (g1);
\vertex [above left = 0.45 cm and 0.2 cm of g1] (v1p);
\vertex [left = 0.875cm of v1p] (fu1);
\vertex [below left = 0.45 cm and 0.2 cm of g1] (v2p);
\vertex [left=0.875cm of v2p] (fu2);
\vertex [above right = 0.625 cm and 0.625 cm of i2] (l1);
\vertex [below right = 0.625 cm and 0.625 cm of i2] (l2);
\vertex [above right=1.5 cm of g1] (v1);
\vertex [below right=1.5cm of g1] (v2);
\diagram* {
(d1) -- [ultra thick, fermion] (fu1),
(fu2) -- [ultra thick, fermion ] (d2),
(fu1) -- [ultra thick,fermion, quarter right, looseness=0.95] (fu2),
(i2) -- [ultra thick,  scalar, half left, looseness=1.7] (g1),
(g1) -- [ultra thick,scalar, half left, looseness=1.7] (i2),
(v1p) -- [ultra thick, photon] (v1),
(v2p) -- [ultra thick, photon] (v2),
};
\end{feynman}
\node at (0.1,0.7) {$\chi$};
\node at (0.1,-0.7) {$\bar{\chi}$};
\node at (3.7,0.7) {$V$};
\node at (3.7,-0.7) {$V$};
\node at (2.8,0) {$\varphi$};
\node at (0.85,0) {$\psi$};
\end{tikzpicture}}\ +\ \scalebox{0.75}{\begin{tikzpicture}[baseline={([yshift=-.75ex]current bounding box.center)},xshift=12cm]
\begin{feynman}[large]
\vertex (i1);
\vertex [right = 1.25cm of i1] (i2);
\vertex [above left=1.5 cm of i2] (d1);
\vertex [below left=1.5cm of i2] (d2);
\vertex [right= 1.25cm of i2] (g1);
\vertex [above left = 0.45 cm and 0.2 cm of g1] (v1p);
\vertex [left = 0.875cm of v1p] (fu1);
\vertex [below left = 0.45 cm and 0.2 cm of g1] (v2p);
\vertex [left=0.875cm of v2p] (fu2);
\vertex [above right = 0.625 cm and 0.625 cm of i2] (l1);
\vertex [below right = 0.625 cm and 0.625 cm of i2] (l2);
\vertex [above right=1.5 cm of g1] (v1);
\vertex [below right=1.5cm of g1] (v2);
\diagram* {
(d1) -- [ultra thick, fermion] (fu1),
(fu2) -- [ultra thick, fermion ] (d2),
(fu1) -- [ultra thick,fermion, quarter right, looseness=0.95] (fu2),
(i2) -- [ultra thick,  scalar, half left, looseness=1.7] (g1),
(g1) -- [ultra thick,scalar, half left, looseness=1.7] (i2),
(g1) -- [ultra thick, photon] (v1),
(g1) -- [ultra thick, photon] (v2),
};
\end{feynman}
\node at (0.1,0.7) {$\chi$};
\node at (0.1,-0.7) {$\bar{\chi}$};
\node at (3.7,0.7) {$V$};
\node at (3.7,-0.7) {$V$};
\node at (2.1,0.2) {$\varphi$};
\node at (0.85,0) {$\psi$};
\end{tikzpicture}}\ + \ \scalebox{0.75}{\begin{tikzpicture}[baseline={([yshift=-.75ex]current bounding box.center)},xshift=12cm]
\begin{feynman}[large]
\vertex (i1);
\vertex [right = 1.25cm of i1] (i2);
\vertex [above left=1.5 cm of i2] (d1);
\vertex [below left=1.5cm of i2] (d2);
\vertex [right= 1.25cm of i2] (g1);
\vertex [above left = 0.45 cm and 0.2 cm of g1] (v1p);
\vertex [left = 0.875cm of v1p] (fu1);
\vertex [below left = 0.45 cm and 0.2 cm of g1] (v2p);
\vertex [left=0.875cm of v2p] (fu2);
\vertex [above right = 0.625 cm and 0.625 cm of i2] (l1);
\vertex [below right = 0.625 cm and 0.625 cm of i2] (l2);
\vertex [above right=1.5 cm of g1] (v1);
\vertex [below right=1.5cm of g1] (v2);
\diagram* {
(d1) -- [ultra thick, fermion] (fu1),
(fu2) -- [ultra thick, fermion ] (d2),
(fu1) -- [ultra thick, scalar, quarter right, looseness=0.95] (fu2),
(fu1) -- [ultra thick, fermion, quarter left, looseness=0.95] (v1p),
(v1p) -- [ultra thick, fermion, quarter left, looseness=0.95] (v2p),
(v2p) -- [ultra thick, fermion, quarter left, looseness=0.95] (fu2),
(v1p) -- [ultra thick, photon] (v1),
(v2p) -- [ultra thick, photon] (v2),
};
\end{feynman}
\node at (0.1,0.7) {$\chi$};
\node at (0.1,-0.7) {$\bar{\chi}$};
\node at (3.7,0.7) {$V$};
\node at (3.7,-0.7) {$V$};
\node at (2.85,0) {$\psi$};
\node at (0.95,0) {$\varphi$};
\end{tikzpicture}}
\end{align*}
\caption{Representative diagrams for loop-induced dark matter annihilation to electrically neutral SM gauge bosons, $\chi \bar{\chi} \to VV$.}
\end{figure}
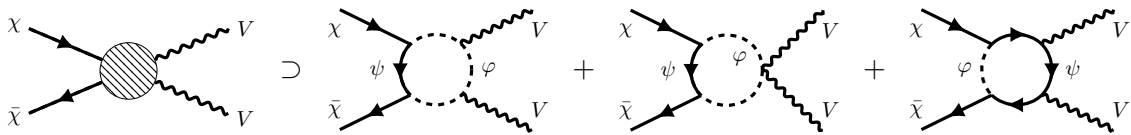

At one loop, the mediators induce effective interactions between pairs of the dark matter and pairs of gluons and hypercharge bosons. 
Representative diagrams are displayed in \hyperref[f1]{Figure 8}.
In the limit in which the dark matter is far lighter than either mediator, the external energy scales are all much smaller than the characteristic momenta inside the loop,
and these interactions map onto a variety of non-renormalizable operators.
The most relevant for the DM relic density, indirect detection, and direct detection are the \emph{Rayleigh operators},
\begin{multline}\label{L0}
\mathcal{L}^{\text{R}}_{\text{eff}} = 
\lambda_{\text{s}}\,\bar{\chi}\chi\, B_{\mu\nu}B^{\mu\nu} + 
\ii \lambda_{\text{p}}\,\bar{\chi}\gamma^5\chi\, B_{\mu\nu}\tilde{B}^{\mu\nu} + 
\kappa_{\text{s}}\,\bar{\chi}\chi \tr G_{\mu\nu}G^{\mu\nu} + 
\ii \kappa_{\text{p}}\,\bar{\chi}\gamma^5\chi \tr G_{\mu\nu} \tilde{G}^{\mu\nu}
\end{multline}
with traces over $\mathrm{SU}(3)_{\text{c}}$ adjoint indices, and the \emph{twist-two} gluon operators,
\begin{align}\label{L2}
    \mathcal{L}^{\text{T}}_{\text{eff}} = 
    \varrho_1\, \bar{\chi}\,\ii\partial^{\{\mu}\gamma^{\nu\}}\chi\, \mathcal{G}^{(2)}_{\mu\nu} + 
    \varrho_2\, \bar{\chi}\,\ii \partial^{\mu}\ii\partial^{\nu} \chi\, \mathcal{G}^{(2)}_{\mu\nu}
\end{align}
with
\begin{align}
\mathcal{G}^{(2)}_{\mu\nu} \equiv \tr \left[G_{\mu}^{\ \,\rho} G_{\rho \nu} + \frac{1}{4}\eta_{\mu\nu}\, G_{\alpha \beta}G^{\alpha\beta}\right],
\end{align}
and where $A^{\{\mu}B^{\nu\}} \equiv A^{\mu}B^{\nu} + A^{\nu}B^{\mu}$. 

We extract the Wilson coefficients $\lambda_{\text{s}}, \lambda_{\text{p}},\kappa_{\text{s}},\kappa_{\text{p}}, \varrho_1,\varrho_2$ in the limit of light DM
by evaluating the one loop amplitudes for $\chi \bar{\chi} \to gg$ and $\chi\bar{\chi} \to BB$ at the point in phase space where the Mandelstam 
invariants\footnote{$s = (k_1+k_2)^2$, $t = (k_1 - p_1)^2$, and $u = (k_1 - p_2)^2$, with $k_1,k_2$ incoming momenta and $p_1,p_2$ outgoing.} 
satisfy $t = u = m_{\chi}^2 - s/2$, and expand in $s$ and $m_{\chi}$ to match onto the 
coefficients of the operators \eqref{L0} and \eqref{L2}. To lowest non-vanishing order,
\begin{align}\label{wilson}
\nonumber \lambda{}_{\text{s}} &= -\frac{\alpha_1}{27\pi}\frac{y_{\chi}^2}{m_{\psi}m_{\varphi}^2},\\
\nonumber \kappa{}_{\text{s}} &= -\frac{5\alpha_3}{96\pi}\frac{y_{\chi}^2}{m_{\psi}m_{\varphi}^2},\\
\nonumber \lambda{}_{\text{p}} &=\tilde{\kappa}{}_{\text{p}} = 0,\\
\nonumber \varrho{}_1 &= \frac{5\alpha_3}{48\pi}\, y_{\chi}^2\, \frac{1}{m_{\varphi}^2 (m_{\varphi}^2-m_{\psi}^2)^2}\left[m_{\varphi}^2 - m_{\psi}^2 + m_{\varphi}^2 \ln \frac{m_{\psi}^2}{m_{\varphi}^2}\right],\\
\text{and}\ \ \ \varrho{}_2 &= \frac{5\alpha_3}{12\pi}\, y_{\chi}^2\,\frac{ m_{\psi}}{(m_{\varphi}-m_{\psi})^5}\left[3(m_{\varphi}^4-m_{\psi}^4) + (m_{\psi}^4 + 4m_{\varphi}^2 m_{\psi}^2 + m_{\psi}^4) \ln \frac{m_{\psi}^2}{m_{\varphi}^2}\right].
\end{align}
The factor of five appearing in the Wilson coefficients for gluon operators can be traced to
the normalization of the generators $\bt{t}^a_{\boldsymbol{6}}$ of the six-dimensional representation of $\mathrm{SU}(3)$ \cite{sextet_catalog,Han_2010}:
\begin{align}\label{dynkin}
\tr \bt{t}^a_{\boldsymbol{6}} \bt{t}^b_{\boldsymbol{6}} = \frac{5}{2}\delta^{ab}.
\end{align}
While we have elected to include the twist-two coefficients for completeness, 
we find their contributions to all relevant processes to be negligible, as expected based on naive power-counting.

\subsection{DM couplings to up-type quarks}

\begin{figure}\label{fquarks}
\begin{align*}
 \scalebox{0.75}{\begin{tikzpicture}[baseline={([yshift=-.75ex]current bounding box.center)},xshift=12cm]
\begin{feynman}[large]
\vertex (i1);
\vertex [right = 1.25cm of i1] (i2);
\vertex [above left=1.5 cm of i2] (d1);
\vertex [below left=1.5cm of i2] (d2);
\vertex [right= 1.25cm of i2] (g1);
\vertex [above left = 0.45 cm and 0.2 cm of g1] (v1p);
\vertex [left = 0.875cm of v1p] (fu1);
\vertex [below left = 0.45 cm and 0.2 cm of g1] (v2p);
\vertex [left=0.875cm of v2p] (fu2);
\vertex [above right = 0.625 cm and 0.625 cm of i2] (l1);
\vertex [below right = 0.625 cm and 0.625 cm of i2] (l2);
\vertex [above right=1.5 cm of g1] (v1);
\vertex [below right=1.5cm of g1] (v2);
\diagram* {
(d1) -- [ultra thick, fermion] (fu1),
(fu2) -- [ultra thick, fermion ] (d2),
(fu1) -- [ultra thick,fermion, quarter right, looseness=0.95] (fu2),
(v1p) -- [ultra thick, fermion, quarter left, looseness=1.1] (v2p),
(fu1) -- [ultra thick,  scalar, quarter left, looseness=0.95] (v1p),
(fu2) -- [ultra thick, scalar, quarter right, looseness=0.95] (v2p),
(v1) -- [ultra thick, fermion] (v1p),
(v2p) -- [ultra thick, fermion] (v2),
};
\end{feynman}
\node at (0.1,0.7) {$\chi$};
\node at (0.1,-0.7) {$\bar{\chi}$};
\node at (3.7,0.75) {$\bar{q}$};
\node at (3.7,-0.75) {$q$};
\node at (1.86, 0.9) {$\varphi$};
\node at (2.9,0.05) {$q^{\text{c}}$};
\node at (0.85,0) {$\psi$};
\end{tikzpicture}}
\end{align*}
\caption{Unique diagram for loop-induced dark matter annihilation to quarks, $\chi \bar{\chi} \to q\bar{q}$. Note the color-conjugated quark $q^{\text{c}}$ in the loop.}
\end{figure}
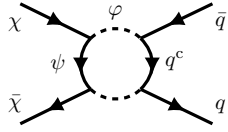

There is an unavoidable coupling induced at one-loop order between the DM and up-type quarks resulting from the diagram of \hyperref[fquarks]{Figure 9}
via the coupling of the scalar mediator in \eqref{L6}. At low energies, the most important effective interactions are the dimension-six operators
\begin{align}\label{LQ}
\mathcal{L}_{\text{eff}}^{\text{Q}} = 
\iota_{II} \left[(\bar{\chi}\gamma^{\mu}\chi)(\bar{u}{}_I\gamma_{\mu}u_I) + (\bar{\chi}\gamma^{\mu}\chi)(\bar{u}{}_I\gamma_{\mu}\gamma^5u_I)\right],\ \ \ I \in \{1,3\},
\end{align}
where we expand the projector $\text{P}_{\text{R}} \propto \boldsymbol{1} + \gamma^5$. 
We extract the Wilson coefficients $\iota_{II}$ in much the same way as the gauge-boson coefficients, expanding in small $s$, $m_{\chi}$ and $m_q$ 
(assuming flavor diagonality in \eqref{L6}, so the final state is always a same-flavor pair) 
and matching onto the amplitude associated with \eqref{LQ}.  To lowest order,
\begin{align}\label{wilsonq}
\iota{}_{II} = \frac{1}{2}\frac{\lambda_{II}^2}{(4\pi)^2}\frac{y_{\chi}^2}{m_{\psi}m_{\varphi}}
\end{align}
for each final state $u_I\bar{u}{}_I$.  While potentially suppressed for small $\lambda_{II}$, these dimension-six operators may nonetheless
dominate over the higher dimensional operators of \eqref{L0}, and can play a relevant role in rate of dark matter annihilation to the $u\bar{u}$ and $t\bar{t}$ final states.

\subsection{Electroweak form factors}
\label{s4.2}

\begin{figure}\label{figmag}
\begin{align*}
\scalebox{0.75}{\begin{tikzpicture}[baseline={([yshift=-.5ex]current bounding box.center)},xshift=12cm]
\begin{feynman}[large]
\vertex [blob] (i1){};
\vertex [below right=0.7 cm and 1.25cm of i1] (i3);
\vertex [below left= 0.7 cm and 1.25cm of i1] (i2);
\vertex [above=1.5cm of i1] (p1);
\diagram* {
(i2) -- [ultra thick, fermion] (i1) -- [ultra thick, fermion] (i3),
(p1) -- [ultra thick, photon] (i1),
};
\end{feynman}
\node at (-1.5,-0.9) {$\chi$};
\node at (1.5,-0.9) {$\chi$};
\node at (-0,1.75) {$\gamma$};
\end{tikzpicture}}\ =\ \scalebox{0.75}{\begin{tikzpicture}[baseline={([yshift=1.5ex]current bounding box.center)},xshift=12cm]
\begin{feynman}[large]
\vertex (i1);
\vertex [above right=1.1cm of i1] (i2);
\vertex [below right=1.1cm of i2] (i3);
\vertex [below left= 1.1cm of i1] (m1);
\vertex [below right=1.1cm of i3] (m2);
\vertex [above=1.1cm of i2] (p1);
\diagram* {
(m1) -- [ultra thick, fermion] (i1),
(i1) -- [ultra thick, scalar] (i2),
(p1) -- [ultra thick, photon] (i2),
(i2) -- [ultra thick, scalar] (i3),
(i3) -- [ultra thick, fermion] (m2),
(i1) -- [ultra thick, fermion, quarter right, looseness=1] (i3),
};
\end{feynman}
\node at (1.07,1.6) {$\gamma$};
\node at (-1,-0.8) {$\chi$};
\node at (2.55,-0.8) {$\chi$};
\node at (0.15,0.55) {$\varphi$};
\node at (0.75,-0.7) {$\psi$};
\end{tikzpicture}}\ +\ \scalebox{0.75}{\begin{tikzpicture}[baseline={([yshift=1.5ex]current bounding box.center)},xshift=12cm]
\begin{feynman}[large]
\vertex (i1);
\vertex [above right=1.1cm of i1] (i2);
\vertex [below right=1.1cm of i2] (i3);
\vertex [below left= 1.1cm of i1] (m1);
\vertex [below right=1.1cm of i3] (m2);
\vertex [above=1.1cm of i2] (p1);
\diagram* {
(m1) -- [ultra thick, fermion] (i1),
(i1) -- [ultra thick, fermion] (i2),
(p1) -- [ultra thick, photon] (i2),
(i2) -- [ultra thick, fermion] (i3),
(i3) -- [ultra thick, fermion] (m2),
(i1) -- [ultra thick, scalar, quarter right, looseness=1] (i3),
};
\end{feynman}
\node at (1.07,1.6) {$\gamma$};
\node at (-1,-0.8) {$\chi$};
\node at (2.55,-0.8) {$\chi$};
\node at (0.1,0.6) {$\psi$};
\node at (0.75,-0.65) {$\varphi$};
\end{tikzpicture}}
\end{align*}
\caption{Diagrams generating a interaction between dark matter and a photon at one loop.}
\end{figure}
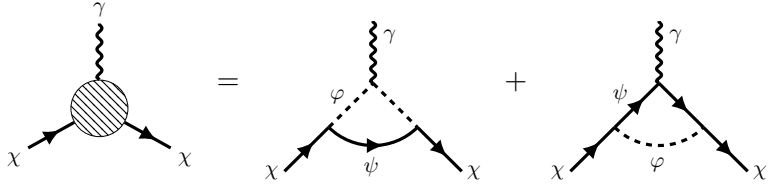

\begin{figure}[t]\label{magplot}
\begin{center}
    \includegraphics[scale=0.7]{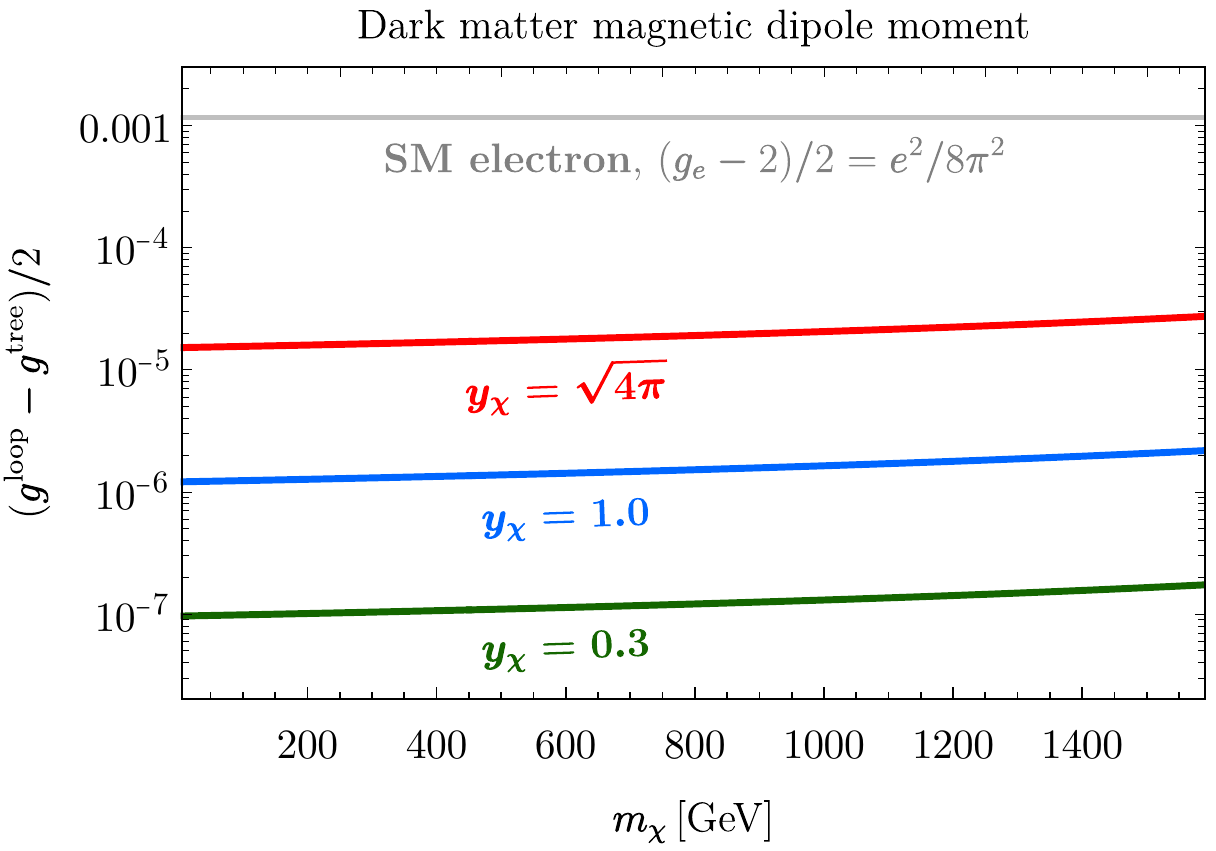}
\end{center}
\caption{One-loop dark matter magnetic dipole moment in benchmark BP1 ($m_{\varphi}=1.15\,\text{TeV}$ and $m_{\psi}=1.6\,\text{TeV}$) for $y_{\chi} \in \{0.3, 1.0,\sqrt{4\pi}\}$. The corresponding result for the electron \cite{PhysRev.73.416,PhysRev.76.790} is shown for reference.}
\end{figure}

At one loop there are contributions to dimension-five and -six electroweak multipole-moment operators
(see \hyperref[figmag]{Figure 10}), which can be important for phenomenology:
\begin{align}
\mathcal{L}_{\text{eff}}^{\text{EM}} =  A_1\, \bar{\chi}\gamma^{\mu}\chi\, \partial^{\nu}B_{\mu\nu} 
+  \frac{1}{4}\, A_2\, \bar{\chi}\sigma^{\mu\nu}\chi\, B_{\mu\nu}
+ A_3\, \bar{\chi}\gamma^{\mu}\gamma^5 \chi\, \partial^{\nu}B_{\mu\nu} 
+  \frac{1}{4}\, A_4\, \ii\bar{\chi}\sigma^{\mu\nu}\gamma^5\chi\,B_{\mu\nu},
\label{mageff}
\end{align}
where $\sigma^{\mu\nu} \equiv \frac{\ii}{2}[\gamma^{\mu},\gamma^{\nu}]$.  The charge-radius operator coefficient $A_1$ and magnetic dipole moment $A_2$ are CP-even
and generic (perhaps even unavoidable) in models with charged mediators and Dirac dark matter.  The
anapole moment $A_3$ and electric dipole moment $A_4$ are CP-odd, and are not generated by minimal implementations of this mediator sector
such as the one we consider.

The Wilson coefficients $A_1$ and $A_2$ are computed for arbitrary momentum transfer in \eqref{ff1} and \eqref{ff2} of \hyperref[a2]{Appendix A}. 
In the limit $m_{\varphi} \approx m_{\psi} \equiv M \gg m_{\chi}$ and $q^2 \approx 4m_{\chi}^2$, relevant for dark matter annihilation with heavy mediators,
\begin{align}\label{Aq4m}
A_1(q^2 \to 4m_{\chi}^2)\equiv A_1^s &= \frac{1}{4}\frac{g_1y_{\chi}^2}{(4\pi)^2}\frac{1}{M^2}\ \ \ \text{and}\ \ \ A_2(q^2\to 4m_{\chi}^2)\equiv A_2^s = \frac{1}{4}\frac{g_1y_{\chi}^2}{(4\pi)^2}\frac{1}{M} .
\end{align}
For non-relativistic scattering with nuclei, the limit $p_1 \approx p_2$ ($q^2 \to 0$) is more appropriate:
\begin{align}\label{Aq0}
\lim_{q^2 \to 0} A_1 (q^2)q^2 = 0\ \ \ \text{and}\ \ \ A_2(q^2 \to 0)\equiv A_2^t = \frac{1}{4}\frac{g_1y_{\chi}^2}{(4\pi)^2}\frac{1}{M}.
\end{align}
Note that $A_k^s$ and $A_2^t$ coincide to $\mathcal{O}(m_{\chi}/M)$ but differ at higher orders.
In \hyperref[magplot]{Figure 11}, we display the exact value of one-loop magnetic dipole moment of the DM,
\begin{align}\label{dipoledef}
 \frac{1}{2}\,g_{\chi} = \frac{1}{e}\,4m_{\chi} \cos \theta_{\text{w}} \times A_2(q^2 \to 0),
 \end{align}
in our first benchmark (BP1, $m_{\varphi}=1.15\,\text{TeV}$ and $m_{\psi}=1.6\,\text{TeV}$) for three illustrative values of $y_{\chi}$. We reiterate that the full results displayed in \hyperref[magplot]{Figure 11} can be found in \hyperref[a2]{Appendix A}.

\section{Dark matter phenomenology}
\label{s5}

In this section, we explore the parameter space of $m_{\chi}$ and $y_{\chi}$, the mass and coupling to the mediators of the dark matter, in light of current constraints
and future prospects for its detection in direct and indirect searches. We also identify the regions of parameter space for which the relic density 
$\Omega_{\chi} h^2$ matches observations, assuming a standard cosmological history during its freeze out.

\subsection{Non-relativistic annihilation}
\label{s5.1}

\begin{figure}\label{treeannfig}
\begin{align*}
    \scalebox{0.75}{\begin{tikzpicture}[baseline={([yshift=-.5ex]current bounding box.center)},xshift=12cm]
\begin{feynman}[large]
\vertex (t1);
\vertex [below=1.5cm of t1] (t2);
\vertex [above left=0.5 cm and 1.1666cm of t1] (i1);
\vertex [right=1.75cm of t1] (f1);
\vertex [above right=0.5cm and 1.1666cm of f1] (p1);
\vertex [below right=0.5cm and 1.1666cm of f1] (p3);
\vertex [below left=0.5cm and 1.1666cm of t2] (i2);
\vertex [right=1.75cm of t2] (f2);
\vertex [below right=0.5 cm and 1.1666cm of f2] (p2);
\vertex [above right=0.5cm and 1.1666cm of f2] (p4);
\diagram* {
(i1) -- [ultra thick, fermion,color=blue] (t1) -- [ultra thick, fermion,color=blue] (t2) -- [ultra thick, fermion,color=blue] (i2),
(f1) -- [ultra thick, scalar] (t1),
(t2) -- [ultra thick, scalar] (f2),
(p3) -- [ultra thick, fermion] (f1),
(f1) -- [ultra thick, fermion] (p1),
(p4) -- [ultra thick, fermion] (f2),
(f2) -- [ultra thick, fermion] (p2),
};
\end{feynman}
\node at (-1.2,0.15) {$\color{blue}\chi$};
\node at (0.9,0.4) {$\varphi^{\dagger}$};
\node at (-1.2,-1.55) {$\color{blue}\bar{\chi}$};
\node at (-0.4,-0.75) {$\color{blue}\psi$};
\node at (0.8,-1.85) {$\varphi$};
\node at (3.26, 0.55) {$q^{\text{c}}$};
\node at (3.2, -0.5) {$\bar{q}$};
\node at (3.26, -0.95) {$\overbar{q^{\text{c}}}$};
\node at (3.2, -2.05) {$q$};
\end{tikzpicture}}
\end{align*}
\caption{Tree-level diagram for dark matter annihilation. $\mathbb{Z}_2$-odd fields are indicated by blue lines and labels.}
\end{figure}
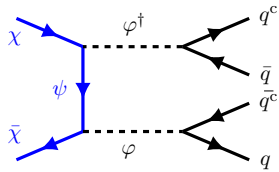

\begin{figure}\label{gammainsertions}
\begin{align*}
\textbf{(a)}\ \ \ \scalebox{0.75}{\begin{tikzpicture}[baseline={([yshift=-.5ex]current bounding box.center)},xshift=12cm]
\begin{feynman}[large]
\vertex (i1){};
\vertex [above left =1.75cm of i1] (x1);
\vertex [below left =1.75cm of i1] (x2);
\vertex [right=2.25cm of i1] (i2);
\vertex [above right=1.75cm of i2] (f1);
\vertex [below right=1.75cm of i2](f2);
\diagram* {
(x1) -- [ultra thick, fermion] (i1) -- [ultra thick, fermion] (x2),
(i1) -- [ultra thick, photon] (i2),
(f1) -- [ultra thick, fermion] (i2) -- [ultra thick, fermion] (f2),
};
\end{feynman}
\node at (-1.4,0.9) {$\chi$};
\node at (-1.4,-0.9) {$\bar{\chi}$};
\node at (3.65,0.9) {$\bar{f}$};
\node at (3.65,-0.9) {$f$};
\node at (1.15, 0.4) {$\gamma,Z$};
\node at (0,0) [circle,draw=black,line width = 0.5mm, fill=gray,inner sep=3.5pt]{};
\end{tikzpicture}}\ \ \ \ \ \ \ \ \ \ \ \ \ \ \ \ \ \ \textbf{(b)}\ \ \ \scalebox{0.75}{\begin{tikzpicture}[baseline={([yshift=-.5ex]current bounding box.center)},xshift=12cm]
\begin{feynman}[large]
\vertex (t1);
\vertex [below=2cm of t1] (t2);
\vertex [above left=0.5 cm and 1.1666cm of t1] (i1);
\vertex [right=1.75cm of t1] (f1);
\vertex [above right=0.5cm and 1.1666cm of t1] (p1);
\vertex [below right=0.5cm and 1.1666cm of t2] (p3);
\vertex [below left=0.5cm and 1.1666cm of t2] (i2);
\vertex [right=1.75cm of t2] (f2);
\vertex [below right=0.5 cm and 1.1666cm of f2] (p2);
\vertex [above right=0.5cm and 1.1666cm of f2] (p4);
\diagram* {
(i1) -- [ultra thick, fermion] (t1) -- [ultra thick, photon] (p1),
(t1) -- [ultra thick, fermion] (t2),
(t2) -- [ultra thick, fermion] (i2),
(t2) -- [ultra thick, photon] (p3),
};
\end{feynman}
\node at (-1.2,0.1) {$\chi$};
\node at (-1.2,-2.1) {$\bar{\chi}$};
\node at (-0.4,-1) {$\chi$};
\node at (1.3, 0.1) {$\gamma,Z$};
\node at (1.3,-2) {$\gamma,Z$};
\node at (0,0) [circle,draw=black,line width = 0.5mm, fill=gray,inner sep=3.5pt]{};
\node at (0,-2) [circle,draw=black,line width = 0.5mm, fill=gray,inner sep=3.5pt]{};
\end{tikzpicture}}
\end{align*}
\caption{Loop-level contributions to dark matter annihilation into \textbf{(a)} SM fermions $f$ (depending on the mediating boson), and \textbf{(b)} annihilation directly to pairs of
bosons.}
\end{figure}
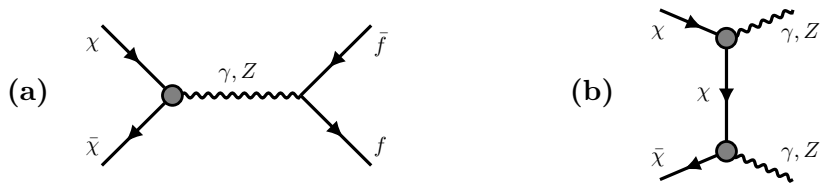

Dark matter annihilation is a key process that controls both its freeze out in the early Universe and potential indirect signals.
At tree level, pairs of dark matter particles can annihilate into pairs of the scalar mediator, which will predominantly be on-shell if $m_\varphi < m_\chi$. The scalar mediators each result in a pair of right-chiral up-type quarks, leading to a four-quark final state as displayed in \hyperref[treeannfig]{Figure 12}. 
These tree-level process(es) dominate annihilation
for $m_{\chi} \gtrsim m_{\varphi}$, and also for a sizable range of $m_{\chi}$ below $m_{\varphi}$.
Below the $\varphi$ threshold, contributions to annihilation from loop-level processes enabled by the interactions discussed in \hyperref[s4]{Section 4} can be very significant,
and allow for annihilation into a pair of fermion or a pair of gauge bosons $\gamma\gamma$, $\gamma Z$, $ZZ$, and $gg$ (see representative diagrams in
\hyperref[gammainsertions]{Figure 13}).

\renewcommand\arraystretch{1.8}
\begin{table}[t]
    \centering
    \begin{tabular}{| c|| c|}
        \toprule
        \hline
       Channel $X$ & $\langle \sigma v_{\chi}\rangle(\chi\bar{\chi} \to X)$\\
       \hline
       \hline
       \rule{0pt}{3.5ex}$f\bar{f}$, $f\in\{\ell,\,q\!\neq \!\{u,t\}\}$ & $\dfrac{17\alpha_Q}{16c_{\text{w}}^2}\, N_{\text{c}}Q_f^2\, [\mathcal{F}(A_1^s, A_2^s)]^2$\\[1ex]
       \hline
       \rule{0pt}{3.5ex}$q_I\bar{q}_I$, $q_I \in \{u,t\}$ & $\dfrac{17\alpha_Q}{16c_{\text{w}}^2}\, N_{\text{c}} Q_{f}^2\, [\mathcal{F}(A_1^s, A_2^s)]^2 + \dfrac{1}{4\pi c_{\text{w}}}\, N_{\text{c}}\,[\mathcal{H}( A_1^s, A_2^s,\iota_{II})]^2$ \\[1ex]
       \hline
       \rule{0pt}{3.5ex}$\nu_{\ell}\bar{\nu}{}_{\ell}$ & $\dfrac{\alpha_Q}{16c_{\text{w}}^2}\, [\mathcal{F}( A_1^s, A_2^s)]^2$\\[1ex]
       \hline
       \rule{0pt}{3.5ex}$W^+ W^-$, $hZ$ & $2\langle \sigma v_{\chi}\rangle(\chi\bar{\chi} \to \nu_{\ell}\bar{\nu}_{\ell})$\\[1ex]
       \hline
       \rule{0pt}{5ex}$\gamma \gamma$ & $\dfrac{2}{\pi}\,c_{\text{w}}^4 m_{\chi}^2\left(\dfrac{A_2^t}{4}\right)^4$\\[1.5ex]
       \hline
       \rule{0pt}{3.5ex}$\gamma Z$ & $2\left(\dfrac{s_{\text{w}}}{c_{\text{w}}}\right)^2 \langle \sigma v_{\chi}\rangle(\chi\bar{\chi} \to \gamma\gamma)$\\[1ex]
       \hline
       \rule{0pt}{3.5ex}$ZZ$ & $\left(\dfrac{s_{\text{w}}}{c_{\text{w}}}\right)^4 \langle \sigma v_{\chi}\rangle(\chi\bar{\chi} \to \gamma \gamma)$\\[1ex]
       \hline
       \rule{0pt}{3.5ex}$gg$ & $\dfrac{64}{\pi}\,m_{\chi}^4 \kappa_{\text{s}}^2 v_{\chi}^2$\\[1ex]
       \hline
       \rule{0pt}{3.5ex}$qq\bar{q}\bar{q}$, $q \in \{u,t\}$ & See \hyperref[a3]{Appendix B}\\[1ex]
       \hline
       \hline
       \multicolumn{2}{|c|}{\rule{0pt}{3.5ex}$[\mathcal{F}(A_1,A_2)]^2 = \left[\dfrac{A_2}{4} + m_{\chi}\,A_1\right]^2$}\\[1ex]
       \hline
       \multicolumn{2}{|c|}{\rule{0pt}{3.5ex}$[\mathcal{H}(A_1,A_2,\iota_{II})]^2 = 2eQ_f m_{\chi}\,\dfrac{A_2 \iota_{II}}{4} + 2eQ_fm_{\chi}^2\,A_1 \iota_{II} + m_{\chi}^2 \iota_{II}^2$}\\[1ex]
       \hline
       \bottomrule
    \end{tabular}
    \caption{Analytic expressions for leading contributions to $\langle \sigma v \rangle (\chi\bar{\chi} \to X)$ for a variety of SM final states $X$, in the limit of negligible SM masses.}
    \label{t2}
\end{table}
\renewcommand\arraystretch{1.0}

Both the relic density from freeze out and the rate of dark matter annihilation relevant for indirect searches hinge on the annihilation cross sections $\langle \sigma v_{\chi} \rangle(\chi\bar{\chi} \to X)$ into the various possible final states $X$ averaged over the distribution of dark matter velocities $v_\chi$. In both cases, the velocities of interest are typically non-relativistic, $v_\chi \sim 0.1$ ($10^{-3}$--$10^{-5}$) for freeze-out (typical observational targets for indirect searches), and it is sufficient to consider the leading terms in an expansion in $v_\chi$.  Approximate analytic expressions for the leading terms of all of the important annihilation channels, in the limit of heavy dark matter (and mediators) such that the masses of the SM particles in the final state can be neglected, are shown in \hyperref[t2]{Table 3}, and a semi-analytic treatment of tree-level annihilation through a pair of mediators in $qq\bar{q}\bar{q}$ is presented in \hyperref[a3.2]{Appendix B}. These approximate analytical results provide a useful guide to understand the relative importance of various channels, but in all numerical results we reinstate full dependence on SM particle masses.
\begin{figure}\label{RX}
\begin{center}
\includegraphics[scale=0.7]{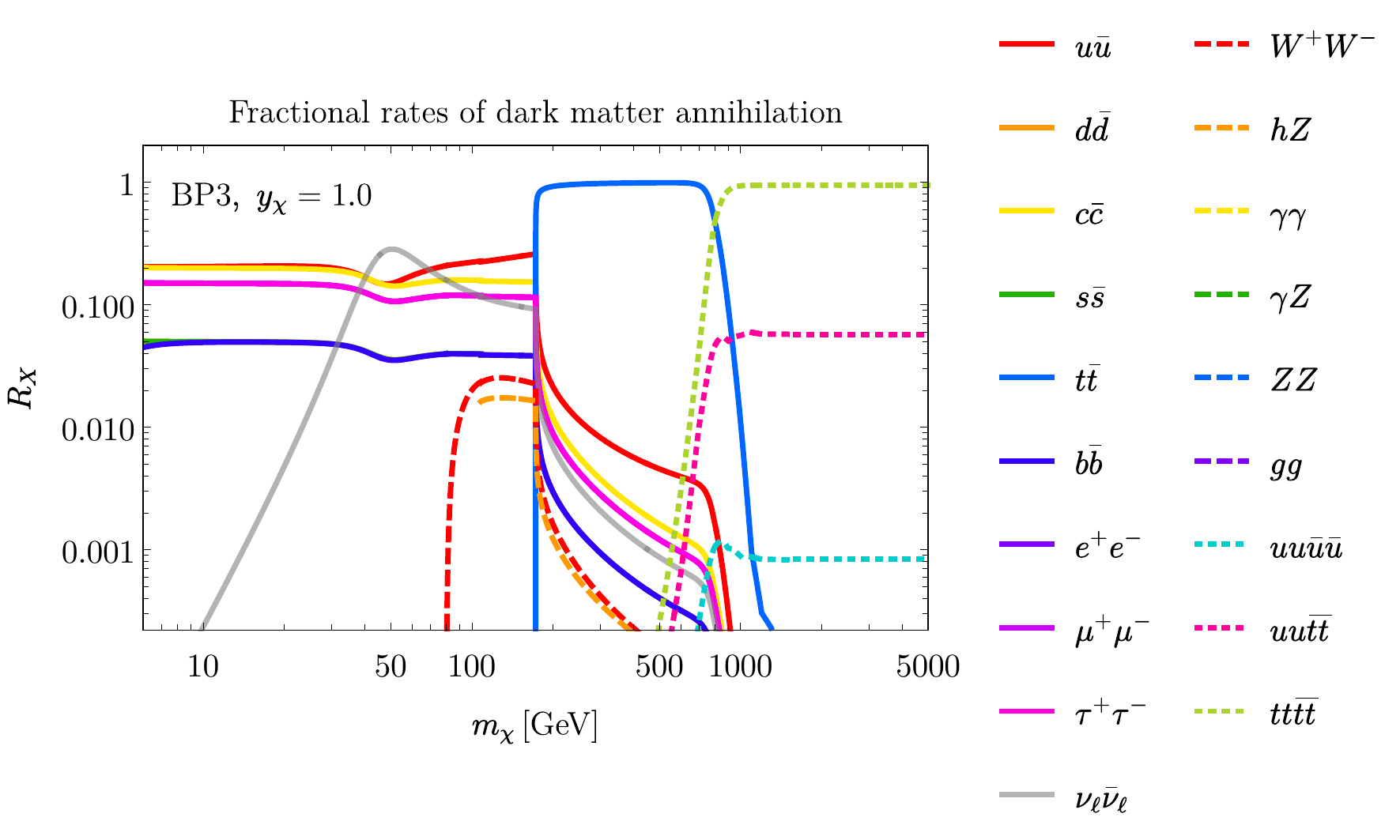}\\[1cm]
\end{center}
\caption{Fraction $R_X$ of total dark matter annihilation rate as a function of $m_{\chi} \leq m_{\psi}$, for nineteen SM final states, in benchmark BP3 ($m_{\varphi}=1.15\,\text{TeV}$, $m_{\psi}=5.0\,\text{TeV}$, $\lambda_{11}=0.125$, and $\lambda_{33}=0.75$). Annihilation fractions are not meaningfully different in BP1 and BP2.}
\end{figure}We first use these results to compute the annihilation fraction,
\begin{align}\label{RXdef}
R_X = \frac{\langle \sigma v_{\chi}\rangle (\chi\bar{\chi} \to X)}{\langle \sigma v_{\chi}\rangle}\ \ \ \text{with}\ \ \ \langle \sigma v_{\chi}\rangle = \sum_X \langle \sigma v_{\chi}\rangle (\chi\bar{\chi} \to X)
\end{align}
for each final state. The annihilation fractions are displayed in \hyperref[RX]{Figure 14} as functions of the DM mass $m_{\chi}$ in the third benchmark BP3 of \hyperref[benchmarktable]{Table 2} with $m_{\chi} \leq m_{\psi} = 5.0\,\text{TeV}$. Every channel considered has the same dependence on $y_{\chi}$, and thus the individual $R_X$ are not sensitive to it. Quark channels typically dominate (pairs of up-type quarks for $m_{\chi} \lesssim m_\varphi /2$ or four quarks above threshold), except for a small region of DM masses around the
$Z$ funnel, where annihilations to neutrinos take over.

\subsubsection{Relic density}\label{s5.2}

\begin{figure}
    \centering
   \includegraphics[scale=0.7]{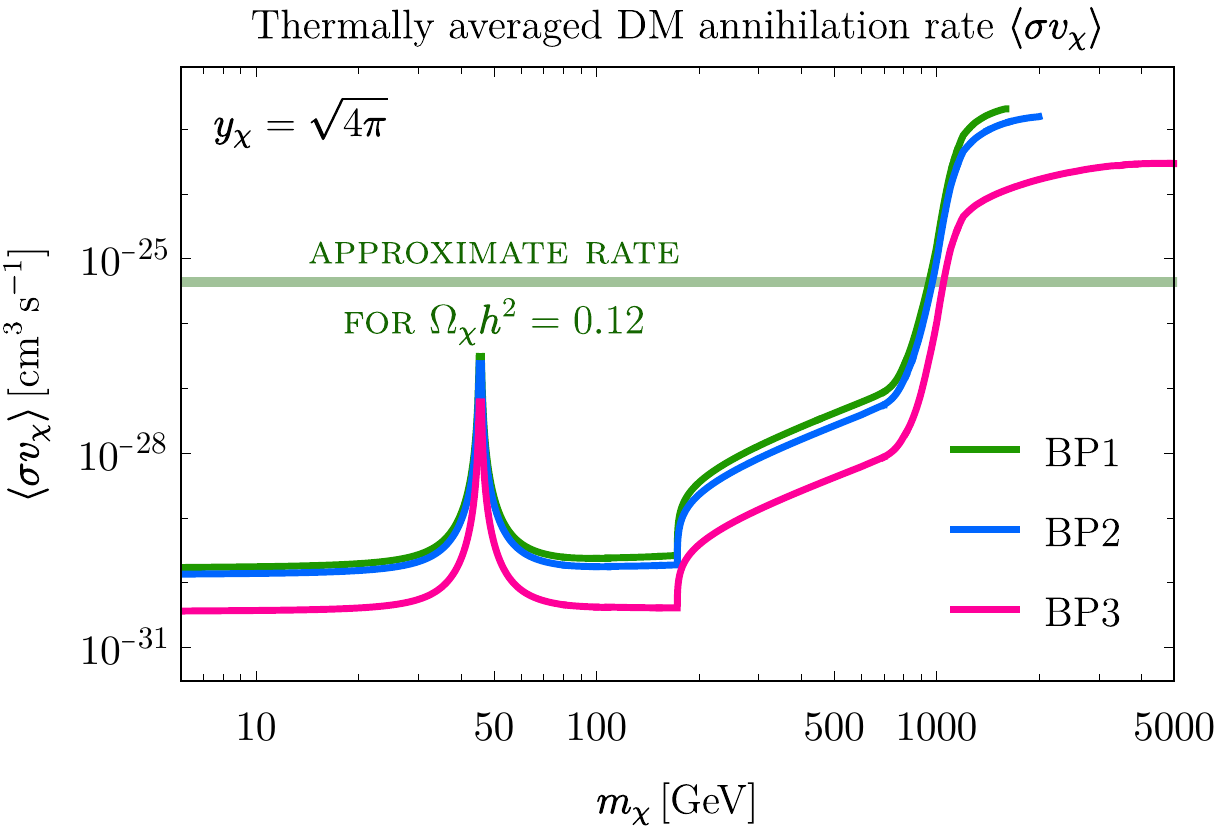}
    \caption{Total dark matter annihilation rate as a function of $m_{\chi} < m_{\psi}$ in all three benchmarks. Results are displayed for $y_{\chi} = \sqrt{4\pi}$, at the upper limit of perturbativity. The line at $\langle \sigma v_{\chi} \rangle \approx 4.4 \times 10^{-26}\,\text{cm}^3\,\text{s}^{-1}$ estimates the annihilation rate producing the observed relic density in a standard cosmological history \cite{sv_2012}.}
    \label{annirate}
\end{figure}

\begin{figure}\label{Ychiplot}
\begin{center}
    \includegraphics[scale=0.7]{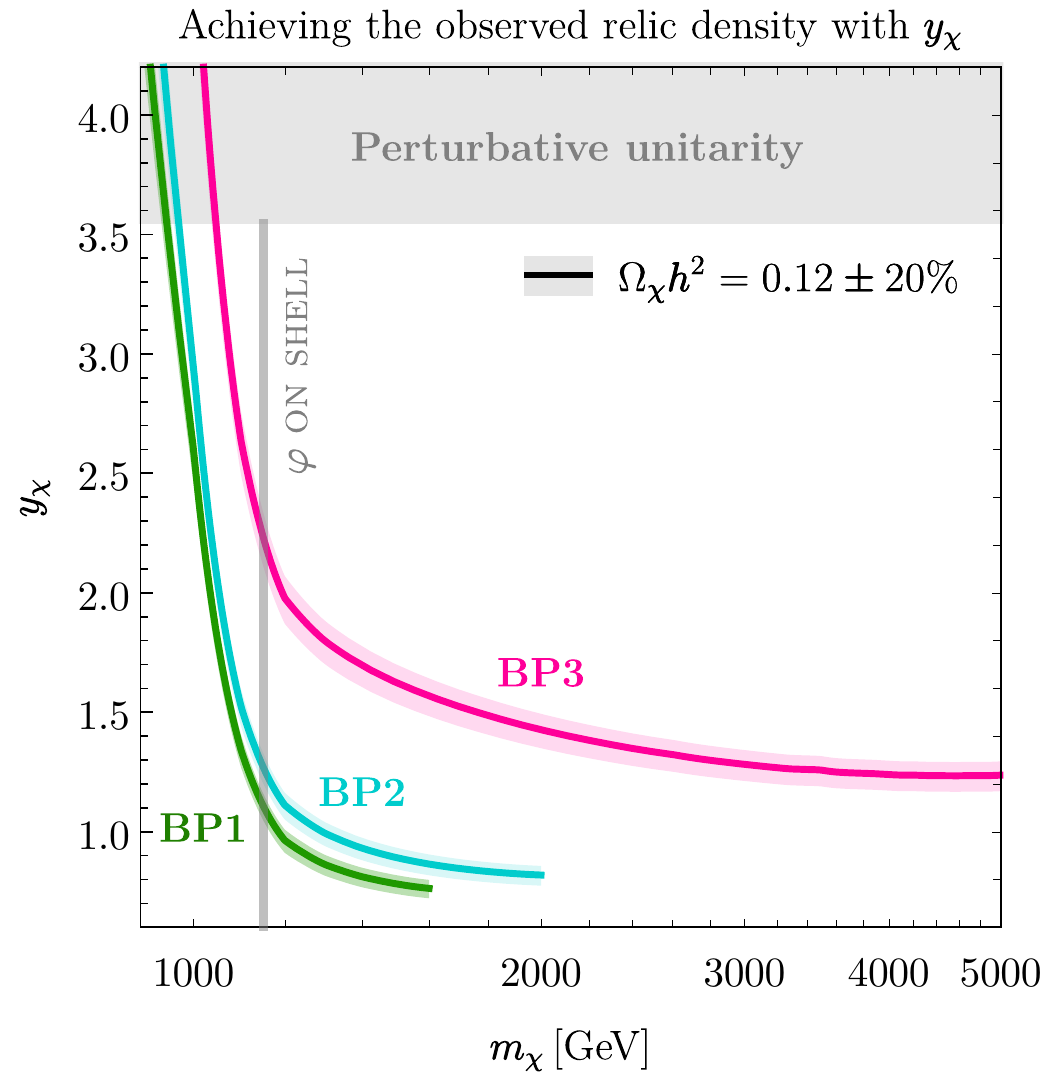}
\end{center}
\caption{Values of $y_{\chi}$ producing the observed relic density as a function of $m_\chi$ in all three benchmarks.}
\end{figure}

If the dark matter $\chi$ evolves according to a standard cosmological history, then the inclusive annihilation rate $\langle \sigma v_{\chi} \rangle$ determines the DM relic abundance. It depends very strongly on the coupling of the dark matter to the mediators $y_{\chi}$.
The annihilation rates in the three benchmarks of \hyperref[benchmarktable]{Table 2} are displayed in \hyperref[annirate]{Figure 15} for $y_{\chi} = \sqrt{4\pi}$,
at the upper limit of perturbativity. This figure compares the benchmark annihilation rates to $\langle \sigma v_{\chi} \rangle \approx 4.4 \times 10^{-26}\,\text{cm}^3\,\text{s}^{-1}$,
the rate producing (approximately) the relic density inferred from fits to the \emph{Planck} data \cite{RD_2020},
\begin{align}\label{relicdensityP}
\Omega_{\chi} h_{\text{Planck}}^2 = 0.120 \pm 0.001,
\end{align}
in a standard cosmology \cite{sv_2012}. \hyperref[annirate]{Figure 15} shows that it is impossible to produce the correct relic density via freeze out through loop-level annihilations. For $m_{\chi} \gtrsim 900\,\text{GeV}$, tree-level annihilations dominate, and a particular value of $y_\chi$ (depending on the mass) will produce the observed relic density. In \hyperref[Ychiplot]{Figure 16}, we show the values of $y_{\chi}$ required to reproduce the observed relic density in each benchmark (assuming a standard cosmology) as a function
of DM mass.

\subsubsection{Indirect searches}\label{s5.2b}

Indirect searches for dark matter seek its visible annihilation products (such as $\gamma$ rays, cosmic rays ($e^+,\bar{p}$), and neutrinos)
originating from local over-densities such as the Galactic Center or Milky Way dwarf spheroidal galaxies (dSphs) \cite{dSphs_1998,dSph_2012}.
As demonstrated in \hyperref[RX]{Figure 14}, annihilation is mostly into quarks and charged leptons for a wide range of $m_{\chi}$, for which
the strongest indirect limits are derived from continuum $\gamma$-ray spectra \cite{ID_2016}.

\begin{figure}
    \centering
    \includegraphics[scale=0.7]{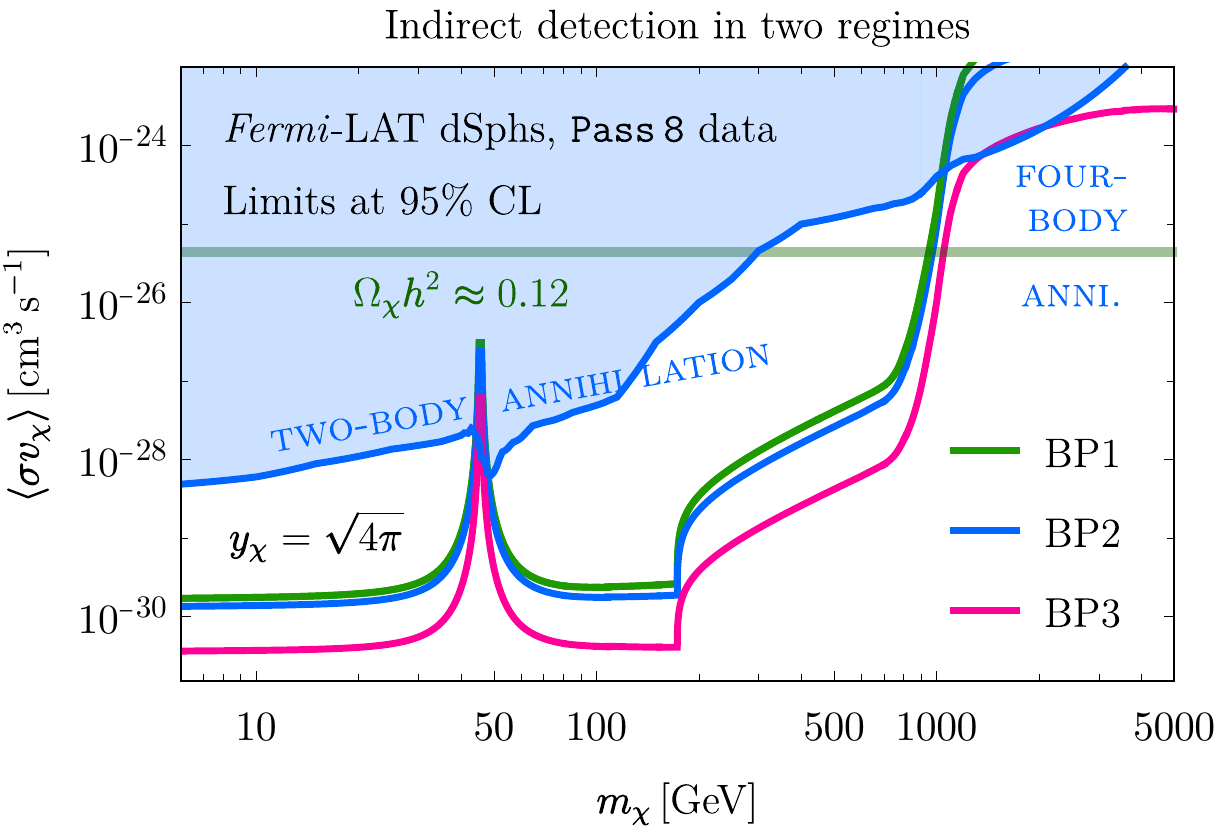}
    \caption{Limits on DM annihilation from $\gamma$-ray emissions from Local Group dwarf spheroidal galaxies (dSphs) observed by the \emph{Fermi}-LAT collaboration. Annihilation rates are displayed for all benchmarks with maximal coupling $y_{\chi} = \sqrt{4\pi}$.}
    \label{darkfluxlims}
\end{figure}

For dark matter lighter than a few TeV, the strongest bounds on continuum $\gamma$ rays from dark matter annihilation are generally from
the null results of \emph{Fermi}-LAT searches for $\gamma$ rays from nearby dSphs \cite{FL_2014,FL_2015}. We impose such limits on our model in two stages, reflecting the two final-state regimes exhibited by our DM candidate. As discussed above, for $m_{\chi} \lesssim 1\,\text{TeV}$, tree-level annihilation through scalar mediators is not kinematically accessible in our benchmarks and all annihilation is at loop level to two-body states. To constrain this regime, we use \textsc{DarkFlux} version 1.0 \cite{Boveia:2022maz}, a tool that combines a set of $2 \to 2$ annihilation processes 
(derived from a UFO model) at leading order and convolves them with the
PPPC4DMID tables \cite{PPPC_2011}, weighted by the appropriate annihilation fractions according to the method developed in \cite{Carpenter:2015xaa,Carpenter:2016thc}, to generate a model-specific $\gamma$-ray spectrum. 
These spectra are fed into the published LAT likelihood functions and statistical methodology to perform a joint-likelihood analysis for the fifteen dSphs considered in the six-year \emph{Fermi}-LAT analysis \cite{FL_2015}, deriving the 95\% CL upper limit on the total annihilation rate $\langle \sigma v_{\chi} \rangle$. 
We implement the one-loop processes as tree-level effective vertices, which we expect to be a good approximation for this mass range. When the dark matter becomes heavier than this, it instead annihilates chiefly to four quarks through on-shell sextet scalars (viz. \hyperref[treeannfig]{Figure 12}). We constrain this $2\to 4$ regime with the help of \textsc{MadDM} version 3.2.1 \cite{Arina:2021gfn}, a plugin for \textsc{MadGraph5\texttt{\textunderscore}aMC@NLO}. \textsc{MadDM} internally uses \textsc{Pythia\,8} version 8.306 to simulate the decays of the color-sextet scalars to quarks and then compute the resulting energy spectra. It then computes the upper limit on $\langle \sigma v_{\chi}\rangle$ in much the same way as \textsc{DarkFlux}. The results of our two-part analysis are displayed in \hyperref[darkfluxlims]{Figure 17}, with the $y_{\chi} = \sqrt{4\pi}$ annihilation rates from \hyperref[annirate]{Figure 15} overlaid to find the limit corresponding to the maximal coupling in each benchmark.
For masses below
the threshold for annihilation into pairs of mediators, the \emph{Fermi}-LAT limits, while strong, only constrain a very narrow window around the $Z$ funnel. For heavier dark matter, meanwhile, \emph{Fermi}-LAT rules out parts of the four-body annihilation region in benchmarks BP1 and BP2 where DM is significantly underabundant, and just barely impinges upon similar space in BP3. Before we move on, we note that there are also important indirect searches for high-energy neutrinos \cite{neu_lim_2021} and monochromatic $\gamma$-ray lines \cite{GC_2015,HESS_2019}. These are both predominantly generated by the electroweak moment operators of \eqref{mageff}, and EFT analyses of such operators \cite{Sigurdson:2004zp,DMEFT_2019,Arina:2020mxo} provide an indication that they are expected to impose limits that are two to three orders of magnitude weaker than the \emph{Fermi}-LAT dSph limits in this mass range.


\subsection{Direct searches}
\label{s5.3}

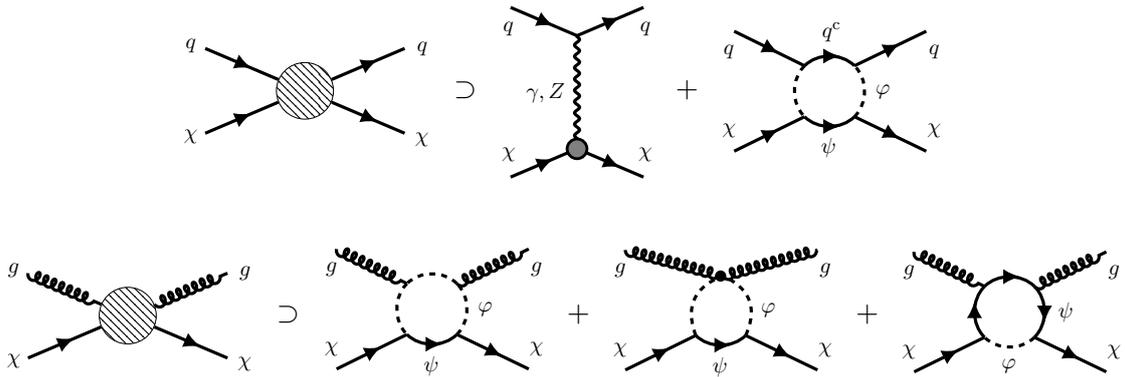
\begin{figure}\label{f3}
\begin{align*}
&\hspace{-5.05cm}\scalebox{0.75}{\begin{tikzpicture}[baseline={([yshift=-.5ex]current bounding box.center)},xshift=12cm]
\begin{feynman}[large]
\vertex [blob] (i1){};
\vertex [above left=0.75 cm and 1.75cm of i1] (d1);
\vertex [below left=0.75 cm and 1.75cm of i1] (d2);
\vertex [above right=0.75 cm and 1.75cm of i1] (v1);
\vertex [below right=0.75cm and 1.75cm of i1] (v2);
\diagram* {
(d1) -- [ultra thick, fermion] (i1),
(d2) -- [ultra thick,fermion] (i1),
(i1) -- [ultra thick, fermion] (v1),
(i1) -- [ultra thick, fermion] (v2),
};
\end{feynman}
\node at (-2,0.8) {$q$};
\node at (-2.0,-0.8) {$\chi$};
\node at (2.06,0.75) {$q$};
\node at (2.05,-0.85) {$\chi$};
\end{tikzpicture}}\ \supset\ \scalebox{0.75}{\begin{tikzpicture}[baseline={([yshift=-.5ex]current bounding box.center)},xshift=12cm]
\begin{feynman}[large]
\vertex (t1);
\vertex [below=2cm of t1] (t2);
\vertex [above left=0.5 cm and 1.1666cm of t1] (i1);
\vertex [right=1.75cm of t1] (f1);
\vertex [above right=0.5cm and 1.1666cm of t1] (p1);
\vertex [below right=0.5cm and 1.1666cm of t2] (p3);
\vertex [below left=0.5cm and 1.1666cm of t2] (i2);
\vertex [right=1.75cm of t2] (f2);
\vertex [below right=0.5 cm and 1.1666cm of f2] (p2);
\vertex [above right=0.5cm and 1.1666cm of f2] (p4);
\diagram* {
(i1) -- [ultra thick, fermion] (t1) -- [ultra thick, fermion] (p1),
(t1) -- [ultra thick, photon] (t2),
(i2) -- [ultra thick, fermion] (t2) -- [ultra thick, fermion] (p3),
};
\end{feynman}
\node at (-1.2,0.13) {$q$};
\node at (-1.2,-2.1) {$\chi$};
\node at (-0.55,-1) {$\gamma,Z$};
\node at (1.2, 0.13) {$q$};
\node at (1.2,-2.1) {$\chi$};
\node at (0,-2) [circle,draw=black,line width = 0.5mm, fill=gray,inner sep=3.5pt]{};
\end{tikzpicture}}\ + \ \scalebox{0.75}{\begin{tikzpicture}[baseline={([yshift=-.75ex]current bounding box.center)},xshift=12cm]
\begin{feynman}[large]
\vertex (i1);
\vertex [right = 1.25cm of i1] (i2);
\vertex [above left=1.5 cm of i2] (d1);
\vertex [below left=1.5cm of i2] (d2);
\vertex [right= 1.25cm of i2] (g1);
\vertex [above left = 0.45 cm and 0.2 cm of g1] (v1p);
\vertex [left = 0.875cm of v1p] (fu1);
\vertex [below left = 0.45 cm and 0.2 cm of g1] (v2p);
\vertex [left=0.875cm of v2p] (fu2);
\vertex [above right = 0.625 cm and 0.625 cm of i2] (l1);
\vertex [below right = 0.625 cm and 0.625 cm of i2] (l2);
\vertex [above right=1.5 cm of g1] (v1);
\vertex [below right=1.5cm of g1] (v2);
\diagram* {
(d1) -- [ultra thick, fermion] (fu1),
(d2) -- [ultra thick, fermion ] (fu2),
(fu2) -- [ultra thick,fermion, quarter right, looseness=0.95] (v2p),
(v2p) -- [ultra thick, scalar, quarter right, looseness=0.95] (v1p),
(fu1) -- [ultra thick, fermion, quarter left, looseness=0.95] (v1p),
(fu1) -- [ultra thick, scalar, quarter right, looseness=0.95] (fu2),
(v1p) -- [ultra thick, fermion] (v1),
(v2p) -- [ultra thick, fermion] (v2),
};
\end{feynman}
\node at (0.1,0.7) {$q$};
\node at (0.1,-0.7) {$\chi$};
\node at (3.7,0.7) {$q$};
\node at (3.7,-0.7) {$\chi$};
\node at (2.8,0) {$\varphi$};
\node at (1.85,-1) {$\psi$};
\node at (1.9, 1.05) {$q^{\text{c}}$};
\end{tikzpicture}}\\[0.75cm]
\scalebox{0.75}{\begin{tikzpicture}[baseline={([yshift=-.5ex]current bounding box.center)},xshift=12cm]
\begin{feynman}[large]
\vertex [blob] (i1){};
\vertex [above left=0.75 cm and 1.75cm of i1] (d1);
\vertex [below left=0.75 cm and 1.75cm of i1] (d2);
\vertex [above right=0.75 cm and 1.75cm of i1] (v1);
\vertex [below right=0.75cm and 1.75cm of i1] (v2);
\diagram* {
(d1) -- [ultra thick, gluon] (i1),
(d2) -- [ultra thick,fermion] (i1),
(i1) -- [ultra thick, gluon] (v1),
(i1) -- [ultra thick, fermion] (v2),
};
\end{feynman}
\node at (-2,0.8) {$g$};
\node at (-2.0,-0.8) {$\chi$};
\node at (2.06,0.75) {$g$};
\node at (2.05,-0.85) {$\chi$};
\end{tikzpicture}}\ \supset\ \scalebox{0.75}{\begin{tikzpicture}[baseline={([yshift=-.75ex]current bounding box.center)},xshift=12cm]
\begin{feynman}[large]
\vertex (i1);
\vertex [right = 1.25cm of i1] (i2);
\vertex [above left=1.5 cm of i2] (d1);
\vertex [below left=1.5cm of i2] (d2);
\vertex [right= 1.25cm of i2] (g1);
\vertex [above left = 0.45 cm and 0.2 cm of g1] (v1p);
\vertex [left = 0.875cm of v1p] (fu1);
\vertex [below left = 0.45 cm and 0.2 cm of g1] (v2p);
\vertex [left=0.875cm of v2p] (fu2);
\vertex [above right = 0.625 cm and 0.625 cm of i2] (l1);
\vertex [below right = 0.625 cm and 0.625 cm of i2] (l2);
\vertex [above right=1.5 cm of g1] (v1);
\vertex [below right=1.5cm of g1] (v2);
\diagram* {
(d1) -- [ultra thick, gluon] (fu1),
(d2) -- [ultra thick, fermion ] (fu2),
(fu2) -- [ultra thick,fermion, quarter right, looseness=0.95] (v2p),
(v2p) -- [ultra thick, scalar, quarter right, looseness=0.95] (v1p),
(v1p) -- [ultra thick, scalar, quarter right, looseness=0.95] (fu1),
(fu1) -- [ultra thick, scalar, quarter right, looseness=0.95] (fu2),
(v1p) -- [ultra thick, gluon] (v1),
(v2p) -- [ultra thick, fermion] (v2),
};
\end{feynman}
\node at (0.1,0.7) {$g$};
\node at (0.1,-0.7) {$\chi$};
\node at (3.7,0.7) {$g$};
\node at (3.7,-0.7) {$\chi$};
\node at (2.8,0) {$\varphi$};
\node at (1.85,-1) {$\psi$};
\end{tikzpicture}}\ &+\ \scalebox{0.75}{\begin{tikzpicture}[baseline={([yshift=-.75ex]current bounding box.center)},xshift=12cm]
\begin{feynman}[large]
\vertex (i1);
\vertex [right = 1.25cm of i1] (i2);
\vertex [above left=1.5 cm of i2] (d1);
\vertex [below left=1.5cm of i2] (d2);
\vertex [right= 1.25cm of i2] (g1);
\vertex [above left = 0.45 cm and 0.2 cm of g1] (v1p);
\vertex [left = 0.875cm of v1p] (fu1);
\vertex [below left = 0.45 cm and 0.2 cm of g1] (v2p);
\vertex [left=0.875cm of v2p] (fu2);
\vertex [above right = 0.625 cm and 0.625 cm of i2] (l1);
\vertex [below right = 0.625 cm and 0.625 cm of i2] (l2);
\vertex [above right=1.5 cm of g1] (v1);
\vertex [below right=1.5cm of g1] (v2);
\vertex [above left= 0.6 and 0.6 cm of g1] (p1);
\vertex [left=0.2 cm of p1] (p2);
\vertex [right=0.2cm of p1] (p3);
\diagram* {
(d1) -- [ultra thick, gluon] (p1) -- [ultra thick, gluon] (v1),
(d2) -- [ultra thick, fermion ] (fu2),
(fu2) -- [ultra thick,fermion, quarter right, looseness=0.95] (v2p),
(fu2) -- [ultra thick, scalar, quarter left, looseness=1.2] (p1),
(v2p) -- [ultra thick, scalar, quarter right, looseness=1.2] (p1),
(v2p) -- [ultra thick, fermion] (v2),
};
\end{feynman}
\node at (0.1,0.7) {$g$};
\node at (0.1,-0.7) {$\chi$};
\node at (3.7,0.7) {$g$};
\node at (3.7,-0.7) {$\chi$};
\node at (2.7,0) {$\varphi$};
\node at (1.85,-1) {$\psi$};
\node at (1.86,0.6)[circle,fill,inner sep=2pt]{};
\end{tikzpicture}}\ + \ \scalebox{0.75}{\begin{tikzpicture}[baseline={([yshift=-.75ex]current bounding box.center)},xshift=12cm]
\begin{feynman}[large]
\vertex (i1);
\vertex [right = 1.25cm of i1] (i2);
\vertex [above left=1.5 cm of i2] (d1);
\vertex [below left=1.5cm of i2] (d2);
\vertex [right= 1.25cm of i2] (g1);
\vertex [above left = 0.45 cm and 0.2 cm of g1] (v1p);
\vertex [left = 0.875cm of v1p] (fu1);
\vertex [below left = 0.45 cm and 0.2 cm of g1] (v2p);
\vertex [left=0.875cm of v2p] (fu2);
\vertex [above right = 0.625 cm and 0.625 cm of i2] (l1);
\vertex [below right = 0.625 cm and 0.625 cm of i2] (l2);
\vertex [above right=1.5 cm of g1] (v1);
\vertex [below right=1.5cm of g1] (v2);
\diagram* {
(d1) -- [ultra thick, gluon] (fu1),
(d2) -- [ultra thick, fermion ] (fu2),
(fu2) -- [ultra thick,fermion, quarter left, looseness=0.95] (fu1),
(fu1) -- [ultra thick, fermion, quarter left, looseness=0.95] (v1p),
(v1p) -- [ultra thick, fermion, quarter left, looseness=0.95] (v2p),
(v2p) -- [ultra thick, scalar, quarter left, looseness=0.95] (fu2),
(v1p) -- [ultra thick, gluon] (v1),
(v2p) -- [ultra thick, fermion] (v2),
};
\end{feynman}
\node at (0.1,0.7) {$g$};
\node at (0.1,-0.7) {$\chi$};
\node at (3.7,0.7) {$g$};
\node at (3.7,-0.7) {$\chi$};
\node at (2.85,0) {$\psi$};
\node at (1.85,-0.95) {$\varphi$};
\end{tikzpicture}}
\end{align*}
\caption{Loop-induced scattering off quarks or gluons, leading to scattering with nuclei. 
The small blob in first diagram denotes an insertion of the electromagnetic moment (one-loop) effective vertex.}
\end{figure}

There are also important searches for ambient dark matter scattering with nuclei via spin-independent (SI) or spin-dependent (SD) interactions. For the low momentum transfer typical
of dark matter local to the Solar System, the SI interactions are typically coherently enhanced for heavy nuclei, and thus are more constraining for 
generic models \cite{DD_EFT_2013_1,RJH_1,DD_EFT_2017}.
The leading diagrams for dark matter to scatter with quarks or gluons are shown in \hyperref[f3]{Figure 18}. Notably, there are no tree-level processes, and the leading
diagrams are at one-loop order.\footnote{Even for loop-suppressed interactions, direct searches often provide relevant constraints \cite{DD_loop_2018,MDMK_2019}.} Because of the small characteristic momentum transfer, higher-order terms in the EFT are typically very subdominant, with the most important one typically being the dimension-five magnetic dipole moment encapsulated by $A_2$.

\begin{figure}
    \centering
    \includegraphics[scale=0.7]{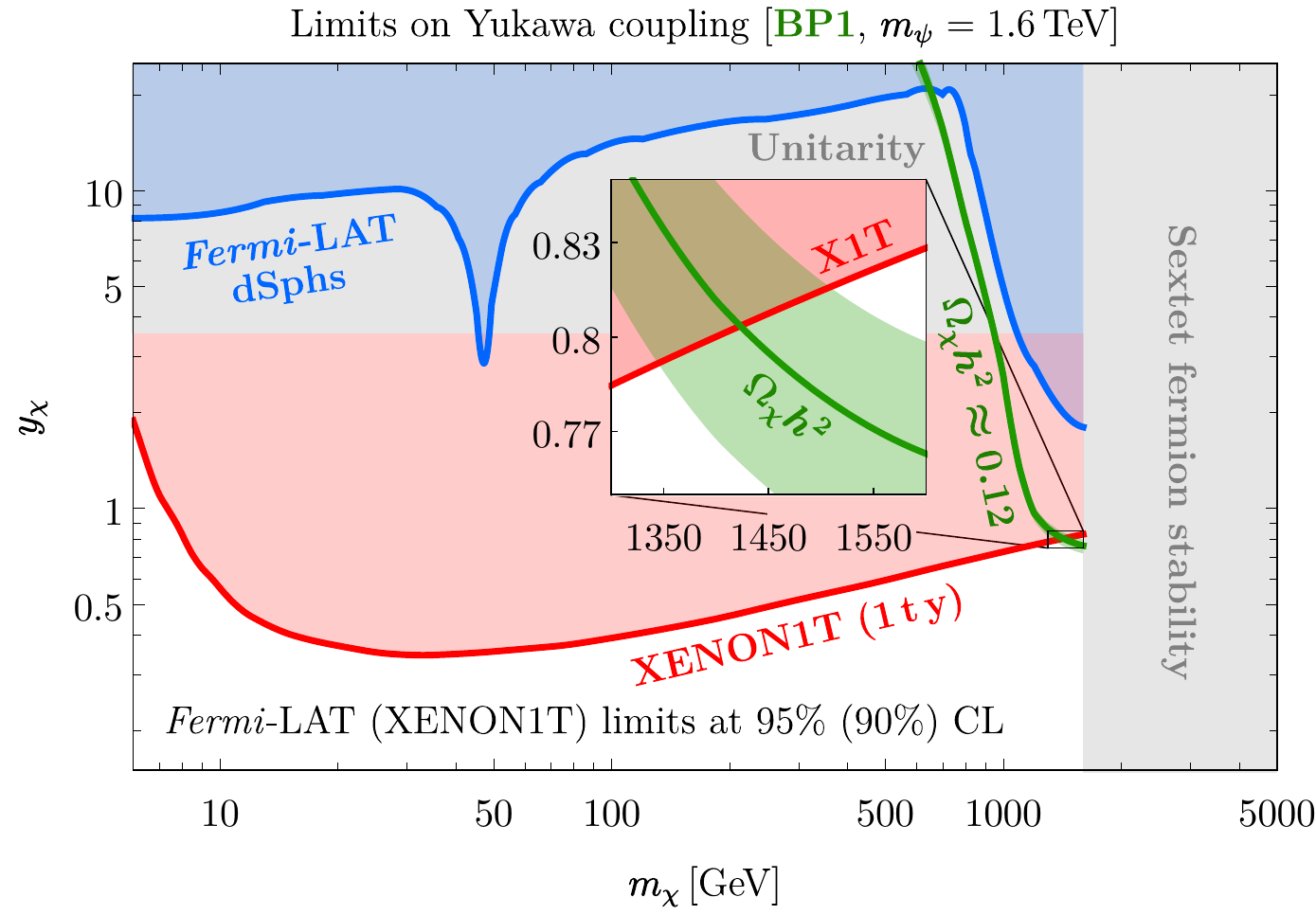}\vspace*{0.75cm}
    \includegraphics[scale=0.7]{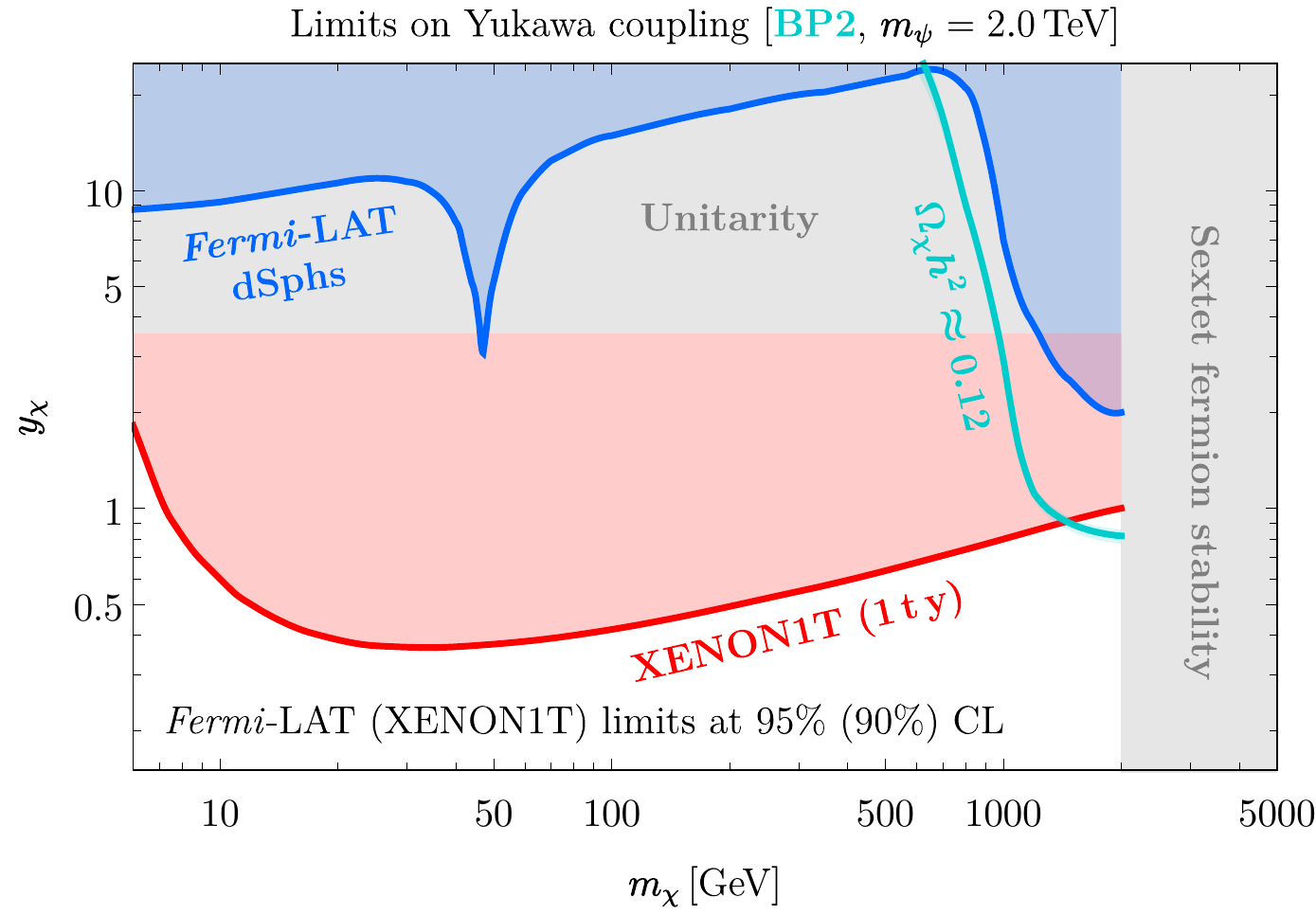}\vspace*{0.75cm}
    \includegraphics[scale=0.7]{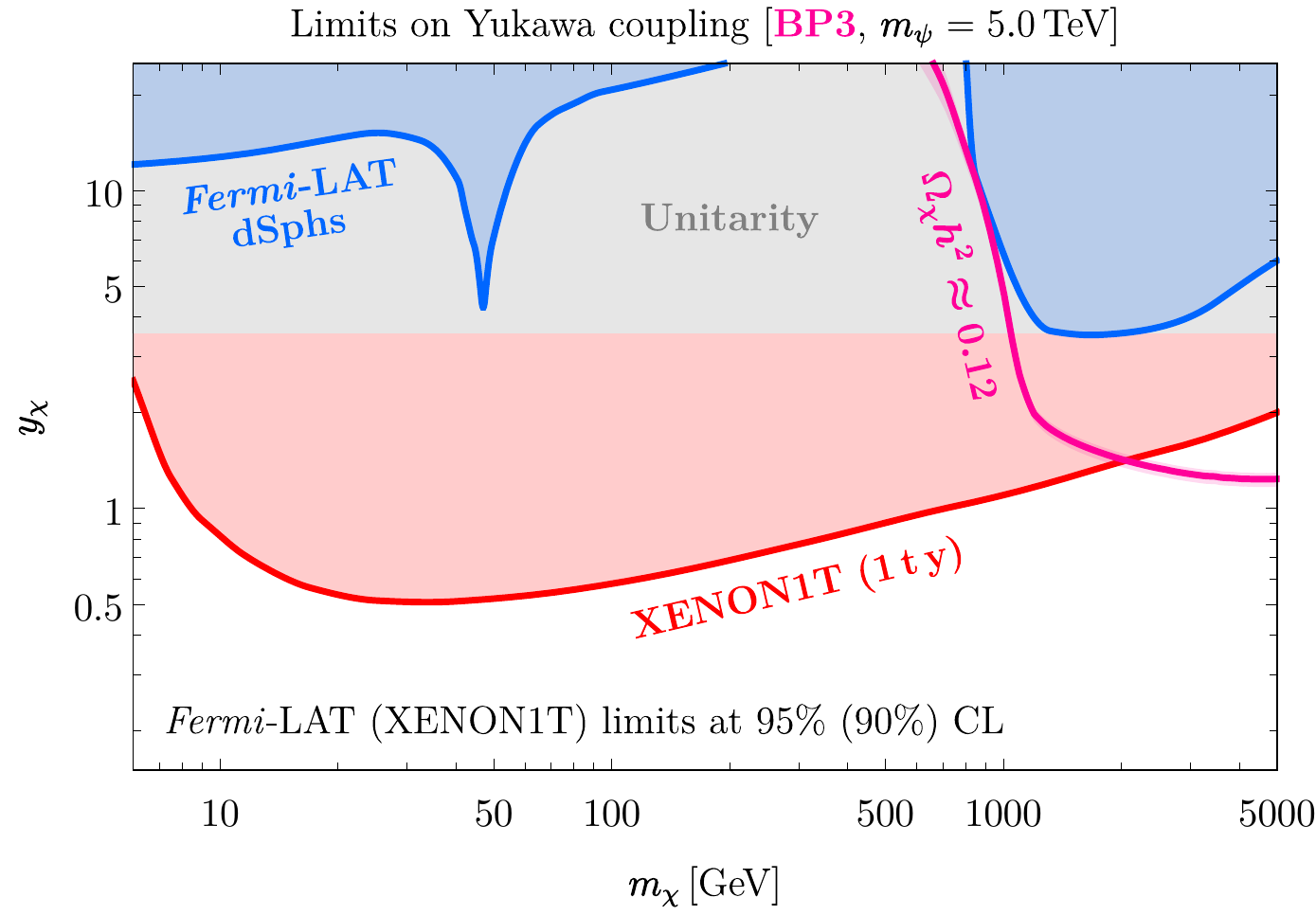}
    \caption{Limits on $y_\chi$ as a function of $m_\chi$ in each benchmark, including direct-detection bounds from XENON1T and the \emph{Fermi}-LAT limits from dSphs.}
    \label{directdetect}
\end{figure}

Currently, the best direct-detection limits derive from the 1 ton-year exposure of XENON1T \cite{X1T_2017,X1T_2018}, which in the absence of a significant excess over background
excludes spin-independent cross sections of dark matter with nucleons
as low as $\sigma_{\text{SI}} = 4.1 \times 10^{-47}\,\text{cm}^2$ for $m_{\text{DM}} = 30\,\text{GeV}$ at $90\%$ CL. These bounds have been mapped onto the parameter space of the dark matter mass and its magnetic moment \cite{DMEFT_2019,Arina:2020mxo}, and we adopt these limits and translate them into the parameter space of $(m_\chi, y_\chi)$, for each of our three benchmark scenarios,
in \hyperref[directdetect]{Figure 19}.
This figure also shows the \emph{Fermi}-LAT dwarf spheroidal limits and the curves indicating the values of $y_\chi$ that result in the observed relic density through freeze-out. The main features of the \emph{Fermi}-LAT limits are the spikes at the $Z$ funnel and the strengthening for $m_{\chi} > m_{\varphi}$ (the four-body annihilation regime). The latter can be understood from \hyperref[darkfluxlims]{Figure 17}: for fixed $y_{\chi}$, $\langle \sigma v_{\chi}\rangle$ increases much faster than the \emph{Fermi}-LAT limits on the annihilation cross section, so the $\gamma$-ray limits on $y_{\chi}$ effectively become stronger in this region. Meanwhile, XENON1T notably disfavors values of $y_{\chi}$ an order of magnitude smaller than the strongest dSph limits. In benchmark BP1, XENON1T further excludes the necessary $y_\chi$ for the relic density for all masses except for a small sliver roughly $200\,\text{GeV}$ wide, where the dark matter becomes close to degenerate with $m_\psi$ (indicated in the inset figure). Given this narrow range, it is possible that the neglected higher-dimensional operators or higher-order effects could prove decisive. At any rate, this window is expected to be excluded by XENONnT, which is underway at the time of writing and projected to improve upon the sensitivity of XENON1T for $\sigma_{\text{SI}}$ by more than an order of magnitude \cite{XnT_2020}. On the other hand, in benchmarks BP2 and BP3, the heavier sextet fermion grants the dark matter significantly more open parameter space with the correct relic density. These results suggest that XENONnT and other future direct searches will probe scenarios such as ours in the multi-TeV range.

On the other hand, reopening some parameter space for lighter mediators and dark matter could be desirable. More dramatic modifications introducing additional physics, such as introducing a Majorana mass term to split the dark matter into two pseudo-Dirac states (which would only scatter inelastically via the magnetic dipole moment) or introducing new production mechanisms, could prove helpful. Alternatively, modifying the early cosmology by e.g. introducing a period of early QCD confinement \cite{Ipek:2018lhm,Berger:2020maa} could enhance the chiral interactions with quarks during freeze-out, allowing for the observed relic density to be realized for smaller $y_\chi$.
\section{Conclusions}
\label{s6}

In this work we have explored a renormalizable model in which dark matter communicates with the Standard Model through a pair of 
mediators in the six-dimensional (sextet) representation of the Standard Model $\mathrm{SU}(3)_{\text{c}}$. 
This model prohibits tree-level couplings of DM pairs to pairs of SM particles, thus \emph{frustrating} the dark matter in its attempts to communicate with the Standard Model, but it generates myriad such couplings at one-loop order. 
It boasts rich phenomenology relevant both for the Large Hadron Collider (LHC) and for independent searches for dark matter. 
We have thoroughly explored its parameter space --- which consists of the DM mass, its coupling to the mediators, the mediator masses, 
and the couplings of the $\mathbb{Z}_2$-even color-sextet mediator to SM quarks --- in several benchmark scenarios via a number of up-to-date terrestrial and astrophysical experiments.

LHC searches for dijet resonances and dijet pairs place limits on sextet-quark couplings of $\mathcal{O}(10^{-1})$ and on the scalar mass around the TeV scale, whereas
searches for events with multiple jets and significant missing transverse energy constrain the $\mathbb{Z}_2$-odd sextet fermion in combination with the dark matter itself.
DM annihilation through a variety of channels determines the parameter space resulting in the correct relic abundance $\Omega_{\chi}h_{\text{Planck}}^2$ through freeze out in a standard cosmology.
We contrast this parameter space with constraints from indirect and direct searches, and find space supporting $\Omega_{\chi}h_{\text{Planck}}^2$ while surviving all experimental constraints --- though the size of this region varies significantly between benchmark scenarios. In a scenario with a light sextet fermion, only a relatively narrow window close to the threshold for annihilation of the dark matter into pairs of mediators survives the strong constraints from XENON1T (which require $y_{\chi} \lesssim \mathcal{O}(10^{-1})$ across most of the available DM mass range), despite the fact that the leading contributions to scattering with nuclei are one-loop suppressed. On the other hand, scenarios with heavier fermions, which in turn accommodate heavier dark matter, produce much broader regions with viable thermal relics.

Our work highlights the fact that straightforward generalizations of the standard simplified model paradigm, in this case invoking coupling to a pair of mediators that themselves connect to the Standard Model, can produce well motivated dark matter models with dramatically expanded phenomenology. Our specific renormalizable scheme allows the dark matter to annihilate to pairs of virtually every known particle, many at appreciable rates. It further illustrates the broad impact that many searches which by themselves are not
motivated directly by popular theories of dark matter --- e.g., in this work, the large collection of LHC searches for singly and pair-produced color-charged resonances --- can have on our understanding of the viable territories of dark matter theory space. In summary, we have introduced a simple but extendable framework that can support viable dark matter candidates despite introducing a second degree of separation between the dark and visible sectors. This framework can stand on its own if the mediator-SM couplings are renormalizable, or it can be considered a low-energy remnant of a fuller theory if higher-dimensional couplings are invoked (which we leave for possible future work). Most importantly, the particular model we have scrutinized in this work places its DM candidate in parameter space well suited to be probed by multiple experiments currently or imminently underway.

\appendix

\section{Electroweak form factors}
\label{a2}

Here we provide complete results for the loop-induced coupling of dark matter pairs to a single photon or $Z$ boson as discussed in \hyperref[s4]{Section 4}. 
The couplings of dark matter to the weak-hypercharge $B$ boson are written as
\begin{align}
\mathcal{L}_{\text{eff}}^{\text{EM}} = A_1\, \bar{\chi}\gamma^{\mu}\chi\, \partial^{\nu}B_{\mu\nu} 
+ \frac{1}{4}\, A_2\, \bar{\chi}\sigma^{\mu\nu}\chi\, B_{\mu\nu}\ \ \ \text{with}\ \ \ \sigma^{\mu\nu} = \frac{\ii}{2}[\gamma^{\mu},\gamma^{\nu}].
\end{align}
The amplitude associated with these operators can be written as
\begin{align}
    \ii \mathcal{M} = \ii \varepsilon_{\mu}(q)\, \bar{u}(p_2)\Gamma^{\mu}(p_1,p_2)u(p_1),
    \end{align}
    with
   \begin{align}\label{formfactormustvanish}
   \Gamma^{\mu}(p_1,p_2)= A_1(q^2) q^2 \left(\gamma^{\mu} - \frac{1}{q^2}\, \slashed{q}q^{\mu}\right) + 4m_{\chi} A_2(q^2)\, \frac{\ii\sigma^{\mu\nu}}{2m_{\chi}}\, q_{\nu},
\end{align}    
where $p_1,p_2$ are the incoming and outgoing DM momenta, $q=p_2-p_1$ is the momentum transferred to the $B$ boson, and $\varepsilon(q)$ is the $B$ polarization vector. 
We write the second term to evoke the Standard Model lepton magnetic dipole moment form factor; 
for the purposes of our discussion in \hyperref[s4.2]{Section 4.2}, we define the DM magnetic dipole moment at one-loop order as\footnote{Clearly, $g_{\chi}^{\text{tree}}=0$; the well known corresponding expression for the electron is $(g_e^{\text{loop}}-g_e^{\text{tree}})/2=(g_e^{\text{loop}}-2)/2 = e^2/8\pi^2$ \cite{PhysRev.73.416}.}
\begin{align}\label{gchi}
\frac{1}{2}\,(g_{\chi}^{\text{loop}}-g_{\chi}^{\text{tree}}) = \frac{1}{2}\,g_{\chi}^{\text{loop}} = \frac{1}{e}\, 4m_{\chi} \cos \theta_{\text{w}} \times A_2(q^2 \to 0)
\end{align}
with $\theta_{\text{w}}$ the weak mixing angle.\footnote{A factor of $\cos \theta_{\text{w}}$ relates the amplitude with a photon to the $B$-boson amplitude.} 

We write the coefficients $A_1, A_2$ in terms of the scalar two- and three-point Passarino-Veltman functions \cite{Passarino:1979pv}
\begin{align}\label{PVfunc}
\nonumber B_0(p^2;m_1^2,m_2^2) & \equiv \int \frac{\d^d \ell}{\ii \pi^{d/2}}\frac{1}{[\ell^2 - m_1^2][(\ell-p)^2 - m_2^2]}\\
    \text{and}\ \ \ C_0(p_1^2,(p_1+p_2)^2,p_2^2;m_1^2,m_2^2,m_3^2) & \equiv \begin{multlined}[t][0cm]\\ \! \! \! \! \! \! \! \! \! \! \! \! \! \! \! \! \! \! \! \! \! \! \! \! \! \! \! \! \! \! \! \int \frac{\d^4 \ell}{\ii \pi^{2}} \frac{1}{[\ell^2-m_1^2][(\ell+p_1)^2-m_2^2][(\ell-p_2)^2-m_3^2]}.\end{multlined}
\end{align}
In particular, we obtain
\begin{multline}\label{ff1}
A_1(q^2) = \frac{1}{2}\frac{g_1 y_{\chi}^2}{(4\pi)^2}\frac{1}{q^2(q^2-4m_{\chi}^2)^2}\bigg[ \!-\!2q^2(q^2-4m_{\chi}^2)+ 12m_{\chi}^2 q^2 B_0(m_{\chi}^2;m_{\varphi}^2,m_{\psi}^2)\\+ \mathcal{B}_1^{(1)}(q^2;m_{\chi},m_{\varphi},m_{\psi}) B_0(q^2;m_{\psi}^2,m_{\psi}^2) - \mathcal{B}_2^{(1)}(q^2;m_{\chi},m_{\varphi},m_{\psi}) B_0(q^2;m_{\varphi}^2,m_{\varphi}^2)\\ - 2 \mathcal{C}_1^{(1)}(q^2;m_{\chi},m_{\varphi},m_{\psi}) C_0(m_{\chi}^2,m_{\chi}^2,q^2;m_{\varphi}^2,m_{\psi}^2,m_{\varphi}^2)\\ -2\mathcal{C}_2^{(1)}(q^2;m_{\chi},m_{\varphi},m_{\psi})C_0(m_{\chi}^2,m_{\chi}^2,q^2;m_{\psi}^2,m_{\varphi}^2,m_{\psi}^2)\bigg]
\end{multline}
and
\begin{multline}\label{ff2}
A_2(q^2) = \frac{1}{2}\frac{g_1 y_{\chi}^2}{(4\pi)^2}\frac{1}{m_{\chi}(q^2-4m_{\chi}^2)^2}\bigg[ \!-\!2m_{\chi}^2(q^2-4m_{\chi}^2)+ 2m_{\chi}^2 (q^2+2m_{\chi}^2) B_0(m_{\chi}^2;m_{\varphi}^2,m_{\psi}^2)\\- \mathcal{B}_1^{(2)}(q^2;m_{\chi},m_{\varphi},m_{\psi}) B_0(q^2;m_{\psi}^2,m_{\psi}^2) - \mathcal{B}_2^{(2)}(q^2;m_{\chi},m_{\varphi},m_{\psi}) B_0(q^2;m_{\varphi}^2,m_{\varphi}^2)\\ + 2 \mathcal{C}_1^{(2)}(q^2;m_{\chi},m_{\varphi},m_{\psi}) C_0(m_{\chi}^2,m_{\chi}^2,q^2;m_{\varphi}^2,m_{\psi}^2,m_{\varphi}^2)\\ +2\mathcal{C}_2^{(2)}(q^2;m_{\chi},m_{\varphi},m_{\psi})C_0(m_{\chi}^2,m_{\chi}^2,q^2;m_{\psi}^2,m_{\varphi}^2,m_{\psi}^2)\bigg]
\end{multline}
with
\begin{align}\label{pieces}
\nonumber \mathcal{B}_1^{(1)}(q^2;m_{\chi},m_{\varphi},m_{\psi}) &= 16m_{\chi}^2[(m_{\chi}+m_{\psi})^2 - m_{\varphi}^2] - 2q^2(7m_{\chi}^2 + m_{\varphi}^2 + 4m_{\chi}m_{\psi} - m_{\psi}^2) + q^4,\\
\nonumber \mathcal{B}_2^{(1)}(q^2;m_{\chi},m_{\varphi},m_{\psi}) &= 16m_{\chi}^2[(m_{\chi}+m_{\psi})^2-m_{\varphi}^2] - 2q^2(m_{\chi}^2 + m_{\varphi}^2 + 4m_{\chi}m_{\psi} - m_{\psi}^2) + q^4,\\
\nonumber \mathcal{C}_1^{(1)}(q^2;m_{\chi},m_{\varphi},m_{\psi}) &= \begin{multlined}[t][10cm] -\!8m_{\chi}^2(m_{\chi}+m_{\psi}+m_{\varphi})(m_{\chi}+m_{\psi}-m_{\varphi})(m_{\chi}^2 + m_{\varphi}^2-m_{\psi}^2)\\ + [m_{\varphi}^4 - 2m_{\varphi}^2(3m_{\chi}^2 - 2m_{\chi}m_{\psi}+m_{\psi}^2)\\ + (m_{\chi}+m_{\psi})(5m_{\chi}^3 + 7m_{\chi}^2 m_{\psi} - 5m_{\chi}m_{\psi}^2 + m_{\psi}^3)]q^2\\ - m_{\psi}(2m_{\chi}-m_{\psi})q^4,
\end{multlined}\\
\mathcal{C}_2^{(1)}(q^2;m_{\chi},m_{\varphi},m_{\psi}) &= \begin{multlined}[t][10cm]
8m_{\chi}^2(m_{\chi}+m_{\psi}+m_{\varphi})(m_{\chi}+m_{\psi}-m_{\varphi}) (m_{\chi}^2 + m_{\psi}^2 - m_{\varphi}^2)\\ + [m_{\varphi}^4 + 2m_{\varphi}^2(5m_{\chi}^2 + 2m_{\chi}m_{\psi}-m_{\psi}^2)\\ -(m_{\chi}+m_{\psi})^2(3m_{\chi}^2 + 6m_{\chi}m_{\psi}-m_{\psi}^2)]q^2\\ + (m_{\chi}+m_{\psi})^2q^4
\end{multlined}
\end{align}
and
\begin{align}\label{pieces2}
\nonumber \mathcal{B}_1^{(2)}(q^2;m_{\chi},m_{\varphi},m_{\psi}) &= m_{\chi}[2m_{\chi}(m_{\chi}^2 + 3m_{\varphi}^2 - 4m_{\chi}m_{\psi}-3m_{\psi}^2) + (m_{\chi}+2m_{\psi})q^2],\\
\nonumber \mathcal{B}_2^{(2)}(q^2;m_{\chi},m_{\varphi},m_{\psi}) &= m_{\chi}[2m_{\chi}(m_{\chi}^2-3m_{\varphi}^2 + 4m_{\chi}m_{\psi}+3m_{\psi}^2) + (m_{\chi}-2m_{\psi})q^2],\\
\nonumber \mathcal{C}_1^{(2)}(q^2;m_{\chi},m_{\varphi},m_{\psi}) &= \begin{multlined}[t][10cm]
m_{\chi}^2 (m_{\chi}+m_{\psi}+m_{\varphi})(m_{\chi}+m_{\psi}-m_{\varphi})\\ \times [(m_{\chi}+3m_{\psi})(m_{\chi}-m_{\psi})+3m_{\varphi}^2]\\ - m_{\chi}[m_{\chi}^3 + 3m_{\chi}^2m_{\psi} + 2m_{\chi}m_{\psi}^2 - m_{\psi}^3 - m_{\varphi}^2(m_{\chi}-m_{\psi})]q^2\\ + \frac{1}{2}m_{\chi}m_{\psi}q^4,
\end{multlined}\\
\mathcal{C}_2^{(2)}(q^2;m_{\chi},m_{\varphi},m_{\psi}) &= \begin{multlined}[t][10cm]
-\!3m_{\varphi}^4m_{\chi}^2 +m_{\chi}(m_{\chi}-m_{\psi})(m_{\chi}+m_{\psi})^2(m_{\chi}^2 + 3m_{\chi}m_{\psi})\\ + 2m_{\chi}^2m_{\varphi}^2(m_{\chi}^2 + 2m_{\chi}m_{\psi} + 3m_{\psi}^2)\\ -m_{\chi}[(m_{\chi}-m_{\psi})(m_{\chi}+m_{\psi})^2 + (2m_{\chi}+m_{\psi})m_{\varphi}^2]q^2.
\end{multlined}
\end{align}
The two kinematic limits of interest are $q^2 \to 4m_{\chi}^2$, relevant for $s$-channel annihilation; and $q^2 \to 0$, appropriate for DM-nucleon scattering, 
$t$-channel annihilation to on-shell photons, and the DM magnetic dipole moment. 
In Sections \hyperref[s4]{4} and \hyperref[s5]{5}, we present the limiting results to first order in $q^2$ and to all orders in the mediator masses. 
Since $m_{\chi} \ll m_{\varphi},m_{\psi}$ in all of the parameter space in which these loops are important, 
we report the results to third order in an expansion in $m_{\chi}/m_{\varphi} \equiv m_{\chi}/m_{\psi}$. In the first limit, we have 
\begin{align}\label{ff4m}
\nonumber A_1(q^2 \to 4m_{\chi}^2) &= \frac{1}{4}\frac{g_1y_{\chi}^2}{(4\pi)^2}\frac{1}{m_{\chi}^2}\left[\left(\frac{m_{\chi}}{M}\right)^2 + \mathcal{O}((m_{\chi}/M)^4)\right],\\
A_2(q^2 \to 4m_{\chi}^2) &= \frac{1}{4}\frac{g_1 y_{\chi}^2}{(4\pi)^2}\frac{1}{m_{\chi}}\left[\frac{m_{\chi}}{M} +\frac{1}{3}\left(\frac{m_{\chi}}{M}\right)^2 + \frac{1}{2}\left(\frac{m_{\chi}}{M}\right)^3+\mathcal{O}((m_{\chi}/M)^4)\right],
\end{align}
and in the second limit, we find
\begin{align}\label{ff0}
\nonumber A_1(q^2 \to 0) &= \frac{1}{4}\frac{g_1y_{\chi}^2}{(4\pi)^2}\frac{1}{3M^2}\left[1 + \frac{1}{10}\left(\frac{m_{\chi}}{M}\right)^2 + \mathcal{O}((m_{\chi}/M)^4)\right],\\
A_2(q^2 \to 0) &= \frac{1}{4}\frac{g_1y_{\chi}^2}{(4\pi)^2}\frac{1}{m_{\chi}}\left[\frac{m_{\chi}}{M} + \frac{1}{3}\left(\frac{m_{\chi}}{M}\right)^2 + \frac{1}{6}\left(\frac{m_{\chi}}{M}\right)^3 + \mathcal{O}((m_{\chi}/M)^4)\right].
\end{align}
The leading terms of these results appear in the body of the paper as \eqref{Aq4m} and \eqref{Aq0}, where we introduce the shorthand
\begin{align}
    A_k(q^2 \to 4m_{\chi}^2) \equiv A_k^s\ \ \ \text{and}\ \ \ A_k(q^2 \to 0) \equiv A_k^t
\end{align}
for $k=1,2$, which is compact and suggestive of the diagram topology for which each set of limiting results is relevant.
\section{Tree-level four-body DM annihilation rate}
\label{a3}

 In renormalizable models with a $\mathbb{Z}_2$-even color-sextet scalar $\varphi$, the dominant tree level DM annihilation channel is
 to four quarks via a pair of $\varphi$. 
 While the largest cross sections are obtained for $m_{\chi} \geq m_{\varphi}$, for which the 
 DM annihilates to approximately on-shell sextet scalars, 
 there is also a sizable parameter space in which dark matter annihilates to quarks through off-shell sextets,
 $\sigma(\chi\bar{\chi} \to q^{\text{c}}{}_I \bar{q}_J\, \overbar{q^{\text{c}}}{}_K q_L)$
(with $I,J,K,L \in \{1,2,3\}$ labeling quark flavors), at a non-negligible rate.
This appendix contains technical details related to the tree-level calculation of the thermally averaged rate of dark matter annihilation 
$\langle \sigma v_{\chi} \rangle$ in the scenario where the sextet-quark couplings are flavor diagonal.

\subsection{Amplitude and kinematics}
\label{a3.1}


We compute cross sections of the form 
$\sigma(\chi\bar{\chi} \to q_I q_I\, \bar{q}{}_J \bar{q}{}_J)$.\footnote{Keeping track of charge conjugated fields, 
in the flavor-diagonal scenario the final state contains two pairs of identical particles. 
There are factors of two throughout reflecting this fact.} The amplitude for this process
can be written as
\begin{multline}\label{4bodyamp}
\ii \mathcal{M}(\chi\bar{\chi} \to q_I q_I\, \bar{q}{}_J \bar{q}{}_J) = -4\ii y_{\chi}^2\,\lambda_{II} \lambda_{JJ}\, \bt{K}_s{}^{ij} \bt{\bar{K}}{}^s{}_{kl}\\ \times \frac{[\bar{v}(k_2)(p_3+p_4+m_{\chi} + m_{\psi})u(k_1)][\bar{u}(p_1)\mathrm{P}_{\text{R}} v(p_2)][\bar{u}(p_3)\text{P}_{\text{L}}v(p_4)]}{[(k_2+p_3+p_4)^2 - m_{\psi}^2][(p_1+p_2)^2-m_{\varphi}^2 + \ii m_{\varphi}\Gamma_{\varphi}(m_{\varphi}^2)][(p_3+p_4)^2-m_{\varphi}^2+ \ii m_{\varphi}\Gamma_{\varphi}(m_{\varphi}^2)]}.
\end{multline}
In this expression, $i,j,k,l \in \{1,2,3\}$ are $\mathrm{SU}(3)_{\text{c}}$ fundamental indices and 
$s$ is an $\mathrm{SU}(3)_{\text{c}}$ sextet index. We label the (incoming) dark matter momenta as $k_1,k_2$ and the 
(outgoing) quark momenta as $p_1,p_2,p_3,p_4$. 
The propagators of the sextet scalars have been promoted to their full Breit-Wigner forms, 
including the energy-dependent sextet decay width
\begin{align}\label{swidth}
\Gamma_{\varphi}(s') = \sum_{I=1}^3 \frac{\lambda_{II}^2}{8\pi}\frac{1}{m_{\varphi}}\, \beta_{q_I}\, (s'-2m_{q_I}^2)\ \ \ \text{with}\ \ \ \beta_X^2 = 1-4\frac{m_X}{s'}.
\end{align}
$\text{P}_{\text{R}}$ and $\text{P}_{\text{L}}$ are the right- and left-chiral projectors, 
and quark spin indices are implied. The on-shell conditions ($\cong$) for the external particles are
\begin{align}\label{onshellcond}
k_1^2, k_2^2 \cong m_{\chi}^2\ \ \ \text{and}\ \ \ p_1^2, p_2^2 \cong m_{q_I}^2\ \ \ \text{and}\ \ \ p_3^2, p_4^2 \cong m_{q_J}^2.
\end{align}

There are eight kinematic degrees of freedom in a $2 \to 4$ process, evident from the four-body differential Lorentz-invariant phase space:
\begin{align}\label{dLIPS}
\d \Pi_4 = \frac{1}{(2\pi)^{12}}\frac{\d^3 p_1}{2p_1^0}\frac{\d^3 p_2}{2p_2^0}\frac{\d^3 p_3}{2p_3^0} \frac{\d^3 p_4}{2p_4^0}\, (2\pi)^4 \delta^4(k_1+k_2 - p_1 - p_2 - p_3 - p_4).
\end{align}
We find it convenient to use a parametrization in which the incoming momenta are written as
\begin{align}\label{kinlist1}
k_{1,2} &= (k^0, \pm k s_{\theta}, 0, \mp k c_{\theta})\ \ \ \text{with}\ \ \ k^0 = \frac{1}{2}\sqrt{s}\ \ \ \text{and}\ \ \ k = \beta_{\chi}k^0,
\end{align}
with $\beta_{\chi}$ the dark matter velocity, and the outgoing momenta are written as
\begin{align}\label{kinlist2}
\nonumber p_{1,2} &= (\gamma_{12}^0 p_{12}^0 \pm \gamma_{12} p_{12} c_{\theta_{12}}, \pm p_{12}s_{\theta_{12}}c_{\phi_{12}}, \pm p_{12} s_{\theta_{12}} s_{\phi_{12}}, \pm \gamma_{12}^0 p_{12} c_{\theta_{12}} + \gamma_{12} p_{12}^0)\\
\text{and}\ \ \ p_{3,4} &= (\gamma_{34}^0 p_{34}^0 \mp \gamma_{34}p_{34}c_{\theta_{34}}, \pm p_{34} s_{\theta_{34}} c_{\phi_{34}}, \pm p_{34} s_{\theta_{34}} s_{\phi_{34}}, \pm \gamma_{34}^0 p_{34} c_{\theta_{34}} - \gamma_{34}p_{34}^0)
\end{align}
with \cite{Byckling:1971vca}
\begin{multline}\label{kinlist3}
\gamma_{12,34}^0 = \frac{1}{2}\frac{s\pm s_{12} \mp s_{34}}{\sqrt{s_{12,34}s}},\ \ \ p_{ab}^0 = \frac{1}{2}\sqrt{s_{ab}},\\ \gamma_{ab} = \frac{1}{2}\frac{\lambda^{1/2}(s,s_{12},s_{34})}{\sqrt{s_{ab}s}},\ \ \ \text{and}\ \ \ p_{ab} = \frac{1}{2}\frac{\lambda^{1/2}(s_{ab}, m_{q_{I'}}^2,m_{q_{I'}}^2)}{\sqrt{s_{ab}s}},
\end{multline}
where $ab \in \{12,34\}$ and $I'=I$ for $ab=12$ and $J$ for $ab=34$, and where
\begin{align}\label{kallen}
\lambda(a,b,c) = a^2 + b^2 + c^2 - 2(ab+ac+bc)
\end{align}
is the triangle function. 
In these expressions, $s_{\alpha}$ and $c_{\theta}$ denote $\sin \alpha,\,\cos\alpha$ for angle $\alpha$. 
These variables can be interpreted as follows \cite{Muta:1986is}:
\begin{enumerate}
    \item $s_{12} \equiv v_1^2= (p_1+p_2)^2$: invariant mass of $\bar{q}{}_I \bar{q}{}_I$ pair,
    \item $s_{34} \equiv v_2^2 = (p_3+p_4)^2$: invariant mass of $q_J q_J$ pair,
    \item $\theta$: scattering angle between $\tv{v}_1$ and $\tv{k}_2$ in $\chi\bar{\chi}$ rest frame,
    \item $\theta_{12}$: decay angle of $\tv{p}_1$ in $\tv{v}_1$ rest frame,
    \item $\phi_{12}$: azimuthal angle about $\tv{v}_1$,
    \item $\theta_{34}$: decay angle of $\tv{p}_3$ in $\tv{v}_2$ rest frame,
    \item $\phi_{34}$: azimuthal angle about $\tv{v}_2$.
\end{enumerate}
The missing eighth degree of freedom is $\phi$, the azimuthal angle about $\tv{k}_2$, over which we can integrate trivially. 
With this parametrization, the differential Lorentz-invariant phase space \eqref{dLIPS} becomes
\begin{align}\label{dLIPS2}
\d \Pi_4 = \frac{1}{(2\pi)^7}\, \d s_{12}\,\d s_{34}\, \d c_{\theta}\, \d \phi_{12}\, \d c_{\theta_{12}}\, \d \phi_{34}\, \d c_{\theta_{34}}\, \mathcal{T}(s; s_{12},s_{34}),
\end{align}
where $\mathcal{T}(s;s_{12},s_{34})$ is given by
\begin{align}\label{Ttri}
    \mathcal{T}(s;s_{12},s_{34}) = \frac{\lambda^{1/2}(s,s_{12},s_{34})}{8s}\frac{\lambda^{1/2}(s_{12},m_{q_I}^2,m_{q_I}^2)}{8s_{12}}\frac{\lambda^{1/2}(s_{34},m_{q_J}^2,m_{q_J}^2)}{8s_{34}}.
\end{align}
When integrating over the complete phase space, the limits of integration are
\begin{align}\label{lims}
\nonumber s_{12} &\in [4m_{q_I}^2,(\sqrt{s}-2m_{q_J}^2)^2],\\
\nonumber s_{34} &\in [4m_{q_J}^2,(\sqrt{s}-\sqrt{s_{12}})^2],\\
\nonumber c_{\theta} &\in [-1,1],\\
\nonumber \phi_{ab} &\in [0,2\pi],\\
\text{and}\ \ \ c_{\theta_{ab}} &\in [-1,1].
\end{align}

\subsection{Cross section}
\label{a3.2}

The squared amplitude can then be written as
\begin{multline}\label{ampsquared}
\langle |\mathcal{M}(\chi\bar{\chi} \to q_I q_I\, \bar{q}{}_J \bar{q}{}_J|^2\rangle \cong -\frac{1}{4}\times {\color{DarkGreen}6} \times [16y_{\chi}^4(\lambda_{II}\lambda_{JJ})^2] \times \frac{1}{\mathcal{D}}\\ \times 8(p_1\doot p_2)(p_3 \doot p_4) \left[ 2s(p_3 \doot p_4 + m_{q_J}^2) - (s-4m_{\chi}^2)(m_{\chi} + m_{\psi})^2 + \mathcal{F} \right],
\end{multline}
with (suppressing the argument of the sextet scalar width)
\begin{align}
 \nonumber   \mathcal{F} &= 4m_{\chi}(m_{\chi}+m_{\psi})(k_1-k_2)\doot(p_3+p_4) - 8[k_1\doot(p_3+p_4)][k_2\doot(p_3+p_4)]\\
    \text{and}\ \ \ \mathcal{D} &= \begin{multlined}[t][10cm] [m_{\chi}^2 - m_{\psi}^2 - 2m_{q_J}^2 + 2p_3 \doot p_4 - 2k_2 \doot (p_3+p_4)]^2\\ \times [2p_1 \doot p_2 + 2m_{q_I}^2 - m_{\varphi}^2 + \ii m_{\varphi}\Gamma_{\varphi}][2p_1 \doot p_2 + 2m_{q_I}^2 - m_{\varphi}^2 - \ii m_{\varphi}\Gamma_{\varphi}]\\ \times [2p_3 \doot p_4 + 2m_{q_J}^2 - m_{\varphi}^2 + \ii m_{\varphi}\Gamma_{\varphi}][2p_3 \doot p_4 + 2m_{q_J}^2 - m_{\varphi}^2 - \ii m_{\varphi}\Gamma_{\varphi}].\end{multlined}
\end{align}
These expressions are the result of a sum over quark colors and spins, and the prefactor of $1/4$ reflects an average over DM spins. 
The {\color{DarkGreen}color factor} is evaluated according to
\begin{align*}
    |\mathcal{M}(\chi\bar{\chi} \to q_I q_I\, \bar{q}{}_J \bar{q}{}_J)|^2 \propto \bt{K}_t{}^{lk}\bt{\bar{K}}{}^t{}_{ji}\bt{K}_s{}^{ij}\bt{\bar{K}}{}^s{}_{kl} = \left|\frac{1}{2}(\delta^i_{\ l}\delta^j_{\ k} + \delta^i_{\ k}\delta^j_{\ l}) \right|^2 = 6,
\end{align*}
where in passing to the second step we have invoked the completeness relation \cite{Han_2010} for a conventional normalization of this set of 
Clebsch-Gordan coefficients. 
Expressed in terms of \eqref{kinlist1}--\eqref{kinlist3}, \eqref{ampsquared} yields a function of $s$ and only three other kinematic variables:
\begin{multline}\label{ampsquaredinv}
\langle |\mathcal{M}(\chi\bar{\chi} \to q_I q_I\, \bar{q}{}_J \bar{q}{}_J)|^2\rangle(s;s_{12},s_{34},c_{\theta}) = -24y_{\chi}^4(\lambda_{II}\lambda_{JJ})^2\\ \times 2(s_{12}-2m_{q_I}^2)(s_{34}-2m_{q_J}^2) \times \frac{\mathcal{F}'}{\mathcal{D}'}
\end{multline}
with
\begin{multline}
\mathcal{F}' = - 8m_{\chi}(m_{\chi} + m_{\psi})\beta_{\chi}c_{\theta}\, s [s^2 - 2s(s_{12}+s_{34}) + (s_{12} - s_{34})^2]^{1/2}\\ + c_{\theta}^2\,(s-4m_{\chi}^2)[s^2 - 2s(s_{12}+s_{34})+(s_{12}-s_{34})^2]\\ - s[4m_{\chi}^2(s-4m_{\psi}^2) + 8m_{\chi}m_{\psi}s - 32 m_{\chi}^3 m_{\psi} - 16m_{\chi}^4 + 4m_{\psi}^2 s - 2s_{34}(s+s_{12}) + (s-s_{12})^2 + s_{34}^2]
\end{multline}
and
\begin{multline}
    \mathcal{D}' = s[(s_{12}-m_{\varphi}^2)^2 + m_{\varphi}^2 \Gamma_{\varphi}^2][(s_{34}-m_{\varphi}^2)^2 + m_{\varphi}^2 \Gamma_{\varphi}^2]\\ \times \left[-\beta_{\chi} c_{\theta}\, [s^2 - 2s(s_{12}+s_{34}) + (s_{12} - s_{34})^2]^{1/2} - s + s_{12} + s_{34} + 2(m_{\chi}^2 -m_{\psi}^2)\right]^2.
\end{multline}
The inclusive cross section is
\begin{multline}\label{xsec}
\sigma(\chi\bar{\chi}\to q_I q_I\, \bar{q}{}_J \bar{q}{}_J) = \frac{1}{2}\frac{1}{(2\pi)^7}\frac{1}{[s(s-4m_{\chi}^2)]^{1/2}}\\ \times \int \d s_{12}\,\d s_{34}\, \d c_{\theta}\, \mathcal{T}(s; s_{12},s_{34})\, \langle |\mathcal{M}(\chi\bar{\chi}\to q_I q_I\, \bar{q}{}_J \bar{q}{}_J)|^2\rangle(s;s_{12},s_{34},c_{\theta})
\end{multline}
with $\mathcal{T}(s;s_{12},s_{34})$ as in \eqref{Ttri} and limits of integration given by \eqref{lims}. This expression includes a factor of $1/4$ for two pairs of identical final-state quarks 
and factors of $4\pi$ from the now-trivial angular phase space integrals. The $c_{\theta}$ integral can be done analytically (and we evaluated it in practice to produce faster numerical results) but this intermediate expression is not illuminating. 
The remaining integrals in \eqref{xsec} are performed numerically, and we check these results for several benchmark points against the 
output of \textsc{MadGraph5\texttt{\textunderscore}aMC@NLO} (\textsc{MG5\texttt{\textunderscore}aMC}) version 3.2.0 \cite{MG5,MG5_EW_NLO}, finding excellent agreement.

The thermally averaged cross section of annihilation to a final state $X$ can be written in terms of an integral over center-of-mass energy as \cite{GONDOLO1991145}
\begin{align}\label{sigv}
    \langle \sigma v_{\chi} \rangle(\chi\bar{\chi} \to X) = \frac{1}{8m_{\chi}^4 T K_2^2(m_{\chi}/T)} \int_{4m_{\chi}^2}^{\infty} \d s\, \sqrt{s}(s-4m_{\chi}^2)\, \sigma(\chi \bar{\chi} \to X)\, K_1(\sqrt{s}/T),
\end{align}
where $T$ denotes the temperature at which the annihilation takes place and $K_n(\xi)$ is the $n^{\text{th}}$-order modified Bessel function of the second kind. 
The formula \eqref{sigv} is valid for $T \lesssim 3m_{\chi}$, and is an 
attractive alternative to expanding $\sigma v_{\chi}$ in $v_{\chi}$ in cases where a cross section cannot be expressed analytically. 
The freeze-out temperature typically occupies a relatively narrow range, $T \in [m_{\chi}/25,m_{\chi}/20]$ \cite{fo_2012}; 
for definiteness, we choose $T = m_{\chi}/22$ in our analysis, but our results are not very sensitive to this choice.

\subsection{On-shell limit}
\label{a3.3}

When the dark matter is heavier than the scalar mediator, on-shell mediator production becomes possible (and indeed accounts for the bulk of the cross section). 
For $m_\chi \geq m_\varphi$, the cross section is well-approximated by the $2 \to 2$ process $\chi\bar{\chi} \to \varphi^{\dagger} \varphi$  with
cross section
\begin{align}\label{onshellsig}
\nonumber    \sigma(\chi\bar{\chi} \to \varphi^{\dagger}\varphi) &= -\frac{1}{16\pi}\frac{1}{(\beta_{\chi}s)^2} \int_{t_0}^{t_1} \d t\, \langle |\mathcal{M}(\chi\bar{\chi} \to \varphi^{\dagger}\varphi)|^2\rangle\\
    &= -\frac{1}{8\pi}\frac{y_{\chi}^4}{(\beta_{\chi}s)^2}\,[h(t_1)-h(t_0)],
\end{align}
where
\begin{align}
    h(t) = -\frac{3}{2}\left[t - \frac{[m_{\varphi}^2 - (m_{\chi} + m_{\psi})^2]^2}{t-m_{\psi}^2} + [s+2(m_{\chi} + m_{\psi})^2 - 2m_{\varphi}^2] \ln\, \frac{t-m_{\psi}^2}{m_{\chi}^2}\right]
\end{align}
and
\begin{align}
    t_0,t_1 = m_{\chi}^2 + m_{\varphi}^2 -\frac{1}{2}s(1\mp \beta_{\chi}\beta_{\varphi}).
\end{align}
We find agreement of 15\% or better between the on-shell result \eqref{onshellsig} for $m_{\chi} \geq m_{\varphi}$ 
and the full result \eqref{xsec} that allows the intermediate sextets to go off shell. 
We finally note that the leading-order contribution to the thermally averaged cross section in the on-shell limit can be written as
\begin{align}\label{onshellvchi}
    \langle \sigma v_{\chi} \rangle(\chi\bar{\chi} \to \varphi^{\dagger}\varphi) &= \frac{y_{\chi}^4}{4\pi}\frac{m_{\chi}^2 - m_{\varphi}^2}{(m_{\chi}^2 + m_{\psi}^2 - m_{\varphi}^2)^2}\left(1-\frac{m_{\varphi}^2}{m_{\chi}^2}\right)^{1/2},
\end{align}
demonstrating that the tree-level annihilation processes are not $p$-wave suppressed. 

\acknowledgments
L.M.C. and T.M. are supported in part by the United States Department of Energy under grant DE-SC0011726. T.T. is supported in part by the United States National Science Foundation under grant no. PHY-1915005. We are grateful to Kirtimaan Mohan for discussions about direct detection at one-loop order.

\bibliographystyle{Packages/JHEP}
\bibliography{Bibliography/bibliography.bib}

\end{document}